\documentclass{article}

\usepackage{lipsum} 

\usepackage[sc]{mathpazo} 
\usepackage[T1]{fontenc} 
\usepackage{microtype} 
\usepackage[hmarginratio=1:1,top=25.4mm,bottom=25.4mm,left=25.4mm,right=25.4mm,columnsep=20pt]{geometry}
\usepackage{multicol} 
\usepackage[hang, small,labelfont=bf,up,textfont=it,up]{caption} 
\usepackage{booktabs} 
\usepackage{hyperref} 

\usepackage{lettrine} 
\usepackage{paralist} 

\usepackage{abstract} 

\usepackage{fancyhdr} 
\usepackage{amsmath}
\usepackage{graphicx}
\usepackage{caption}
\usepackage{subcaption}
\usepackage{amsfonts}
\usepackage{rotating}
\usepackage{amsthm}
\usepackage{MnSymbol}
\usepackage{tikz}

\usetikzlibrary{arrows}
\usepackage{wasysym}

\usepackage{comment}

\usepackage{tikz}
\usetikzlibrary{arrows,positioning,automata}
\tikzset{
	>=stealth',
	punkt/.style={
		rectangle,
		rounded corners,
		draw=black, very thick,
		text width=4em,
		minimum height=4em,
		text centered,
        font=\sffamily\LARGE},
	pil/.style={
		->,
		thick,
		shorten <=2pt,
		shorten >=2pt,}

}
\definecolor{davy}{rgb}{.5,.5,.5}

\title{A dynamical systems model of unorganised segregation in two neighbourhoods}

\author{D. J. Haw and S. J. Hogan\thanks{D. J. Haw, School of Public Health, Imperial College London. S. J. Hogan: Department of Engineering Mathematics, University of Bristol, Bristol BS8 1UB}}

\date{}
\begin{document}

\maketitle

\begin{abstract}

We present a complete analysis of the Schelling dynamical system \cite{Haw2018} of two connected neighbourhoods, with or without population reservoirs, for different types of linear and nonlinear tolerance schedules.

We show that stable integration is only possible when the minority is small and combined tolerance is large. Unlike the case of the single neighbourhood, limiting one population does not necessarily produce stable integration and may destroy it. 

We conclude that a growing minority can only remain integrated if the majority increases its own tolerance. Our results show that an integrated single neighbourhood may not remain so when a connecting neighbourhood is created.

\end{abstract}

\section{Introduction}
Segregation in society takes on many forms, occurring not only in ethnicity or religion, but also gender, for example amongst toddlers \cite{MollerSerbin1996} and in professional hierarchies \cite{Clifton2019}. Segregation is seen as so divisive that some have
argued it ``puts the whole idea of a peaceful society with its constitutional and civic
liberties at risk." (cited in \cite[p.4]{PancsVriend2007}), with \texteuro 1.1B allocated to integration efforts by the Netherlands in 2003 (cited in \cite[p.4]{PancsVriend2007}).

Schelling`s seminal work on segregation \cite{Schelling1969, Schelling1971} is well known for its use of agent-based modelling to explain unorganised\footnote{The modern usage would be \textit{self-organised}.} segregation. These papers have been cited thousands of times and inspired innumerable other works. They are also the subject of controversy in terms of precedence \cite{Hegelsmann2017} and applicability \cite{Fossett2006}.

What appears less well known is that, in the same two papers, Schelling introduced another model, the bounded neighbourhood model (BNM) of unorganised segregation. 
In Schelling`s BNM, a neighbourhood is like a district within a city. Within the neighbourhood, every member is concerned about the overall population {\it mixture}, not with any particular configuration. A member moves out if they are not happy with the population mixture. 

Suppose that the population of a single neighbourhood is divided into two types and let $X(t), Y(t) \ge 0$  represent the respective population sizes, as a function of time $t$. 
In this neighbourhood, {\it tolerance limits} are allocated to a given population type via a {\it tolerance
schedule}, as follows\footnote{Tolerance is a measure of how members of one population remain in an area where there is \textit{another} population present. In contrast, homophily (or self-segregation) \cite{Clifton2019} is a measure of how much one population seeks out members of the \textit{same} population.}. The $X$-population tolerance schedule $R_X(X)$ describes the minimum ratio $Y/X$ required in order
for all of the $X$-population to remain in that neighbourhood. 
A similar function $R_Y(Y)$ denotes the tolerance schedule of the $Y$-population. 

Schelling \cite{Schelling1969, Schelling1971} made the following assumptions:
\begin{itemize}
\item[S1.] The neighbourhood is preferred over other locations: populations of either type will enter/remain/leave unless tolerance conditions are violated. 
\item[S2.] The tolerance schedule is neighbourhood-specific.
\item[S3.] Each member of the population is aware of the ratio of population types within the
neighbourhood at the moment the decision is made to enter/remain/leave (perfect information).
\item[S4.] There is no lower bound on tolerance: no population insists on the presence of the
other type.
\item[S5.] Tolerance schedules are monotone decreasing: the more tolerant population members are the first
to enter and the last to leave.
\end{itemize}


We gave the first dynamical systems formulation of Schelling's BNM in \cite{Haw2018}. This work studies the continuous movement of two populations in and out of a single neighbourhood. We presented the first complete quantitative analysis of the model for {\it linear} tolerance schedules. A fully predictive model was derived and each term within the model was associated with a social meaning. Schelling`s qualitative results were recovered and generalised. 

For the case of unlimited population movement, we derived exact formulae for regions in parameter space where stable integrated populations can occur and showed how {\it neighbourhood tipping} can be explained in terms of basins of attraction. When population numbers are limited, we derived exact criteria for the occurrence of integrated populations.

A natural extension of \cite{Haw2018} is to consider multiple neighbourhoods, constructing sets of differential equations that describe the flow of population within and between these neighbourhoods. 
 In this paper, we focus on the case of two populations $X(t), Y(t)$ moving within and between two neighbourhoods. Our work is related to the "two-room model" of segregation studied in \cite{Shin1, Shin2}. 
Our approach differs in that 
we work in continuous time, whereas they work in discrete time.

We structure the paper as follows. 
In Section \ref{sec:case1}, we consider the situation in which both populations are contained solely within the two neighbourhoods, so that any population leaving one neighbourhood must necessarily relocate to the other. We consider linear tolerance schedules, examining cases when the tolerance schedules are the same or different in both neighbourhoods. 

We examine the case when one population has its numbers limited in one area (Section \ref{sec:limit}), and look at how nonlinearity in the tolerance schedules can change outcomes (Section \ref{sec:nonlin}).

We discuss our results in Section \ref{sec:disc}.
We consider the situation where there are reservoirs of both populations, outside the two neighbourhoods. Whilst the total population of each type in the whole system is conserved, populations can enter or exit either neighbourhood without recourse to  the other neighbourhood.  We also examine similarities between the one- and the two-neighbourhood problems and consider what happens when a second neighbourhood is added to the single neighbourhood problem.

Our conclusions are presented in Section \ref{sec:conclusion}.



\section{Linear tolerance schedules}
\label{sec:case1}
\begin{figure}[h!]
\centering
\includegraphics[width=0.4\linewidth]{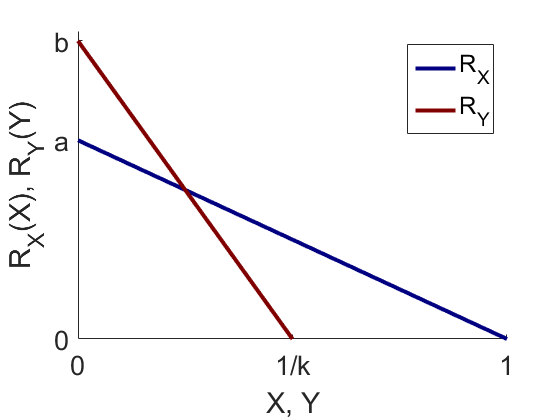}
\caption{Linear tolerance schedules $R_X(X), R_Y(Y)$, as defined in \eqref{eq:tol}.}
\label{fig:tol}
\end{figure}

Recall that for one neighbourhood with two populations $X,Y$, where population members can come and go depending on their tolerance, the dynamical system derived in \cite[Eq. (27)]{Haw2018} is given by
\begin{eqnarray}
\label{eq:orig}
\frac{dX}{dt}&=&X[XR_X(X)-Y],\\
\frac{dY}{dt}&=&Y[YR_Y(Y)-X] \nonumber
\end{eqnarray}
for general tolerance schedules $R_X(X), R_Y(Y)$. This model satisfies the constraints on the $X$-population that $\frac{dX}{dt}\gtrless 0$ when $R_X(X)\gtrless \frac{Y}{X}$: in other words, the $X$-population grows (decays) when the tolerance schedule $R_X(X)$ exceeds (falls short of) $Y/X$ and that $\frac{dX}{dt}=0$ when $X=0$. The model satisfies similar constraints on the $Y$-population.
Schelling's initial example of a tolerance schedule is linear, as shown in Figure \ref{fig:tol}. We set
\begin{eqnarray}
\label{eq:tol}
R_X(X)&=&a(1-X),\\
R_Y(Y)&=&b(1-kY),\nonumber
\end{eqnarray}
where we scale the size of the $X$-population to $1$ and the minority $Y$-population to $1/k$, where $k\geq 1$. With our scalings, the most tolerant member of the $X$-population can abide a $\frac{Y}{X}$ ratio of $a > 0$  and the least tolerant member of the $X$-population can not abide any members of the $Y$-population. Likewise, the most tolerant member of the $Y$-population can abide a $\frac{X}{Y}$ ratio of $b > 0$ and the least tolerant member of the $Y$-population can not abide any members of the $X$-population. 

In the one-neighbourhood problem, it is assumed that outside the neighbourhood, there is a ``place where colour does not matter" \cite{Schelling1971}. Population members can move between this place and the neighbourhood. 

We now consider the case of two neighbourhoods and two populations, where any population leaving one neighbourhood must necessarily enter the other. Let $(X_i,Y_i), \enskip i=1,2$ denote the $(X,Y)$-populations in neighbourhood $i$. So $X_1+X_2=X_{total}$ and $Y_1+Y_2=Y_{total}$ where $X_{total}, Y_{total}$ are both constant. As above, we scale populations such that  $X_{total}=1$ and $Y_{total}=\frac{1}{k}$.  Hence 
\begin{eqnarray}
X_2&=&1-X_1, \label{eq:x1x2}\\
Y_2&=&\frac{1}{k}-Y_1. \label{eq:y1y2}
\end{eqnarray} 

If the neighbourhoods were completely independent of one another, then $(X_i,Y_i), \enskip i=1,2$ would separately satisfy \eqref{eq:orig}.
But the population dynamics in one neighbourhood is affected by movement of population to and from the other neighbourhood. So we have to take that into account in our modelling. To do this we make the additional mild assumption that each member in both neighbourhoods only cares about the population mixture of the neighbourhood that they are in.

Let us consider neighbourhood 1. The dynamics is governed by the following equations.

\begin{eqnarray}
\frac{dX_1}{dt}&=&X_1[X_1R_{X_1}(X_1)-Y_1]-X_2[X_2R_{X_2}(X_2)-Y_2]\label{eq:case1X} \\ 
\frac{dY_1}{dt}&=&Y_1[Y_1R_{Y_1}(Y_1)-X_1]-Y_2[Y_2R_{Y_2}(Y_2)-X_2]
\label{eq:case1Y}
\end{eqnarray}

The first two terms on the right hand side of \eqref{eq:case1X} are the same as those in \eqref{eq:orig} and correspond to the movement of $X$-population, subject to the presence of the $Y$-population, in and out of neighbourhood 1. The third and fourth terms account for the movement of the $X$-population, subject to the presence of the $Y$-population, in and out of neighbourhood 2. The terms on the right hand side of  \eqref{eq:case1Y} can be considered in an analogous way. 

For linear tolerance schedules \eqref{eq:tol} in both neighbourhoods, for $i = 1,2$ we set 
\begin{eqnarray}
\label{eq:toli}
R_{X_i}(X_i)&=&a_i(1-X_i),\\
R_{Y_i}(Y_i)&=&b_i(1-kY_i). \nonumber
\end{eqnarray}
where $a_1 \ne a_2,\,b_1 \ne b_2$ in general.
From \eqref{eq:x1x2} and \eqref{eq:y1y2}, the dynamics in neighbourhood 2 is given simply by
\begin{eqnarray}
\frac{dX_2}{dt}&=&-\frac{dX_1}{dt},\\ 
\frac{dY_2}{dt}&=&-\frac{dY_1}{dt}.
\end{eqnarray}
So we have no need to consider these dynamics explicitly, since they can be obtained directly from the dynamics of neighbourhood 1. 

Substituting \eqref{eq:x1x2}, \eqref{eq:y1y2} into \eqref{eq:case1X} and \eqref{eq:case1Y}, with linear tolerance schedules \eqref{eq:toli}, we obtain the following equations for $(X_1,Y_1)$:

\begin{eqnarray}
\label{eq:linunscaled}
\frac{dX_1}{dt}&=&a_1X_1^2(1-X_1)-X_1Y_1-a_2X_1(1-X_1)^2+(1-X_1)(\frac 1k-Y_1),\\
\frac{dY_1}{dt}&=&b_1Y_1^2(1-kY_1)-X_1Y_1-kb_2Y_1(\frac 1k-Y_1)^2+(1-X_1)(\frac 1k-Y_1).\nonumber
\end{eqnarray}
Note that the coordinate axes $X_1=0,\,Y_1=0$ are no longer nullclines of the full system (unlike in \cite{Haw2018}). 

We can reduce the ensuing algebraic complexity by performing an additional scaling. Define new variables $\hat{t}, \ \hat{Y}_1$ given by 
\begin{eqnarray}
\label{eq:rescale}
\hat{t}=\frac{t}{k},\ \hat{Y}_1=\frac{Y_1}{a_1}
\end{eqnarray}
 and then set 
\begin{eqnarray}
\label{eq:parms}
\alpha=ka_1,\ \beta_1=a_1b_1,\ \beta_2=a_1b_2,\ \gamma=\frac{a_2}{a_1}.
\end{eqnarray}

Now we drop the hats and simplify \eqref{eq:linunscaled} to find:
\begin{eqnarray}
\label{eq:case1nondim}
\frac{dX_1}{dt} & = & (1-X_1)[1-\alpha\gamma X_1 + \alpha(1+\gamma)X_1^2]-\alpha Y_1,\\
a_1\frac{dY_1}{dt} & = & (1-\alpha Y_1)[1-\beta_2Y_1+\alpha(\beta_1+\beta_2)Y_1^2]-X_1. \nonumber
\end{eqnarray}
Equilibria (steady states) in neighbourhood $1$ are given by $(X_1,Y_1) = (X_1^e,Y_1^e)$ where $\frac{dX_1^e}{dt}=\frac{dY_1^e}{dt}=0$. If both $X_1^e \ne 0$ and $Y_1^e \ne 0$, these equilibria correspond to integration. 
We find $(X_1^e,Y_1^e)$ by considering the intersection in $(X_1,Y_1)$-space of the nullclines  of \eqref{eq:case1nondim}, namely solutions of  
\begin{eqnarray}
\alpha Y_1&=&(1-X_1)[1-\alpha\gamma X_1 + \alpha(1+\gamma)X_1^2],\label{sub1}\\
X_1&=&(1-\alpha Y_1)[1-\beta_2Y_1+\alpha(\beta_1+\beta_2)Y_1^2].\label{sub2}
\end{eqnarray}

We will establish conditions under which real positive solutions of \eqref{sub1} and \eqref{sub2} can exist. Then we will find further conditions under which such solutions are stable. 

\subsection{Case I}
We consider the case when the $(X,Y)$-population linear tolerance schedules are identical in both neighbourhoods. So in \eqref{eq:toli}, we set $$a_1=a_2=a,\ b_1=b_2=b,$$ Hence from \eqref{eq:parms} we have
\begin{eqnarray}
\label{eq;parmssymm}
\alpha = ka,\ \beta_1=\beta_2=\beta, \ \gamma=1.
\end{eqnarray} 
Parameters $\alpha$ and $\beta$ are key to what follows. With scaling \eqref{eq:rescale}, both refer to the minority $Y$-population: large/small $\alpha$ corresponds to a small/large minority and large/small $\beta$ refers to a tolerant/intolerant minority.
From \eqref{eq:case1nondim}, the governing equations become
\begin{eqnarray}
\label{eq:case1anondim}
\frac{dX_1}{dt}&=&(1-X_1)[1-\alpha X_1 + 2\alpha X_1^2]-\alpha Y_1,\\
a_1\frac{dY_1}{dt}&=&(1-\alpha Y_1)[1-\beta Y_1+2\alpha\beta Y_1^2]-X_1 \nonumber
\end{eqnarray}

Our aim is to find how many possible equilibria $(X_1^e,Y_1^e)$ exist and then to examine their stability. The points $(X_1^e,Y_1^e)$ correspond to the intersection of the nullclines

\begin{eqnarray}
\alpha Y_1&=&(1-X_1)[1-\alpha X_1 + 2\alpha X_1^2],\label{sub1a}\\
X_1&=&(1-\alpha Y_1)[1-\beta Y_1+2\alpha\beta Y_1^2].\label{sub2a}
\end{eqnarray}


Substitution of \eqref{sub1a} into \eqref{sub2a} gives an equation of the form $$p_9(X_1)=0,$$ where $p_9(X_1)$ is a real polynomial of degree 9 in $X_1$. So there are at most 9 equilibria of \eqref{eq:case1anondim}. 

By inspection,  we have $(X_1^e,Y_1^e)=(1,0), (0,\frac 1{\alpha}),\ (\frac 12,\frac 1{2\alpha})$ for any values of the parameters $\alpha, \beta$. These solutions correspond to: 
\begin{itemize}
    \item[$(X_1^e,Y_1^e)=(1,0)$:] all the $X$-population and none of the $Y$-population in neighbourhood 1; none of the $X$-population and all the $Y$-population in neighbourhood 2.
    \item[$(X_1^e,Y_1^e)=(0,\frac 1{\alpha})$:] none of the $X$-population and all the $Y$-population in neighbourhood 1; all the $X$-population and none of the $Y$-population in neighbourhood 2. 
    \item[$(X_1^e,Y_1^e)=(\frac 12,\frac 1{2\alpha})$:] both $X,Y$-populations evenly split between neighbourhoods 1 and 2. 
\end{itemize}

So we can write 
\begin{eqnarray}
\label{eq:p9}
p_9(X_1)&=&X_1(1-X_1)(1-2X_1)p_6(X_1)
\end{eqnarray}
where 
\begin{eqnarray}
\label{eq:p6}
p_6(X_1) & \equiv & \sum_{i=0}^6a_iX_1^i
\end{eqnarray} 
is a real polynomial of degree 6. It can be shown that $p_6(X_1)=p_6(X_2)=p_6(1-X_1)$, from \eqref{eq:x1x2}. This gives $a_5=-3a_6,\ a_3=5a_6-2a_4,\ a_1=-3a_6+a_4-a_2$ and so the odd coefficients of $p_6(X_1)$ can be written in terms of the even coefficients. We find that
\begin{eqnarray}
\label{eq:ai}
a_0&=&\alpha+\beta+\frac{\beta}{\alpha}\\
a_1&=&-3\beta(\alpha+2) \nonumber \\
a_2&=&\beta(2\alpha^2+15\alpha+6) \nonumber \\
a_3&=&-12\alpha\beta(\alpha+2) \nonumber \\
a_4&=&2\alpha\beta(13\alpha+6) \nonumber \\
a_5&=&-24\alpha^2\beta \nonumber \\
a_6&=&8\alpha^2\beta. \nonumber
\end{eqnarray}


Since both $\alpha,\beta >0$, the signs of $a_i$ in $p_6(X_1)$ alternate. Hence by Descartes' rule of signs, $p_6(X_1)$ can have $0,2,4$ or $6$ positive real roots $(X_1=X_1^e>0)$, depending on parameter values, and no negative real roots. By the symmetry of $p_6(X_1)$, any roots   $X_1^e \in [0,1]$. Hence $p_9(X_1)$ can have $3,5,7$ or $9$ positive real roots, no negative real roots and any roots must lie in the interval $[0,1]$.

Note that we could have chosen to eliminate $X_1$ from \eqref{sub1a} and \eqref{sub2a} to give a ninth order polynomial $q_9(Y_1)$ in $Y_1$. Similar considerations would then apply to the roots of $q_9(Y_1)$ with any roots $Y_1^e \in [0,\frac 1{\alpha}]$. 

\begin{figure}
	\centering
	\begin{subfigure}{.49\textwidth}
		\centering
		\includegraphics[width=\linewidth]{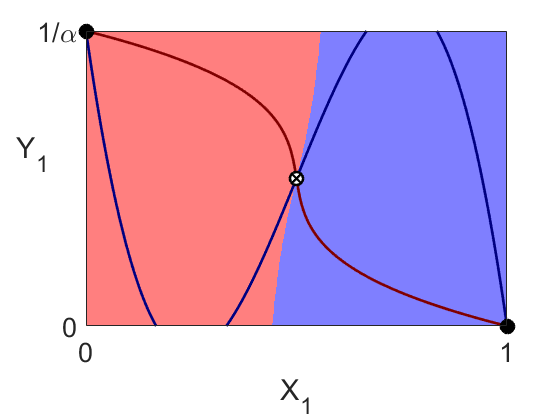}
		\caption{$\alpha=9,\ \beta=16$: $(X_1, Y_1)$ phase space}
		\label{case1aEg1}
	\end{subfigure}
    \begin{subfigure}{.49\textwidth}
		\centering
		\includegraphics[width=\linewidth]{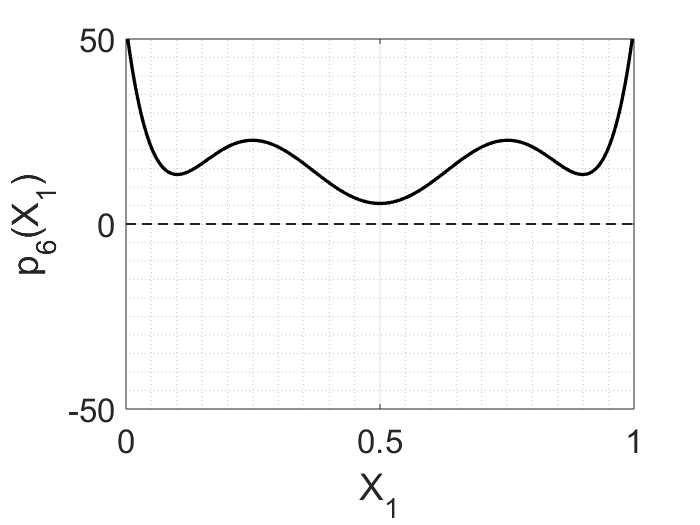}
		\caption{$\alpha=9,\ \beta=16$: $p_6(X_1)$ vs $X_1$}
		\label{case1aEg1p6}
	\end{subfigure}\\
	\begin{subfigure}{.49\textwidth}
		\centering
		\includegraphics[width=\linewidth]{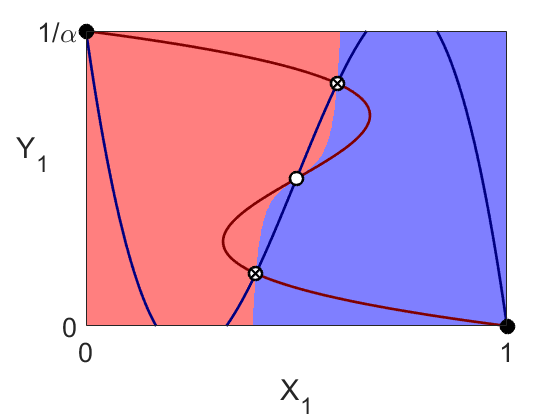}
		\caption{$\alpha=9,\ \beta=40$: $(X_1, Y_1)$ phase space}
		\label{case1aEg2}
	\end{subfigure}
    \begin{subfigure}{.49\textwidth}
		\centering
		\includegraphics[width=\linewidth]{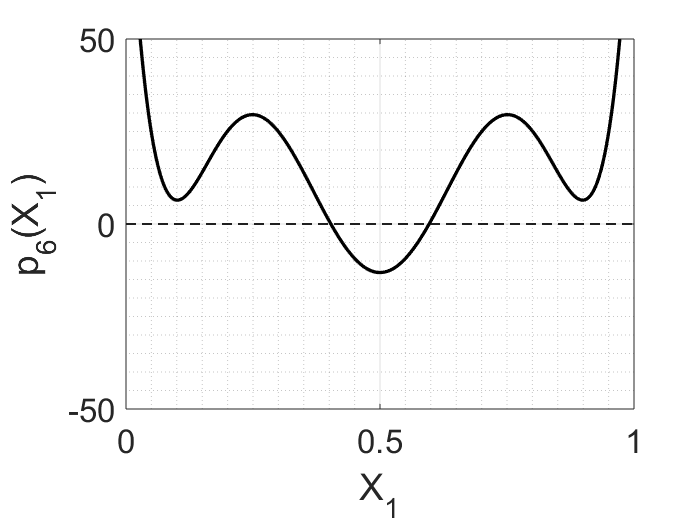}
		\caption{$\alpha=9,\ \beta=40$: $p_6(X_1)$ vs $X_1$}
		\label{case1aEg2p6}
	\end{subfigure}\\
	\begin{subfigure}{.49\textwidth} 
		\centering
		\includegraphics[width=\linewidth]{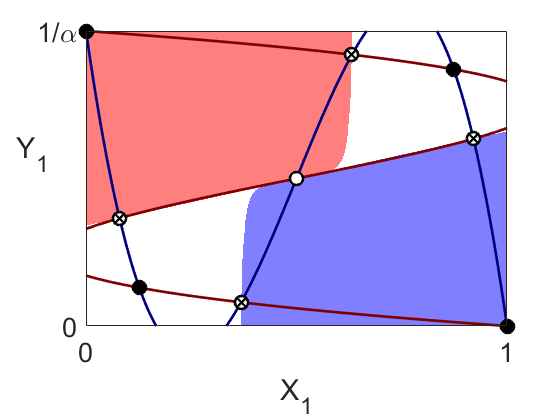}
		\caption{$\alpha=9,\ \beta=80$: $(X_1, Y_1)$ phase space}
		\label{case1aEg3}
    \end{subfigure}
    \begin{subfigure}{.49\textwidth} 
		\centering
		\includegraphics[width=\linewidth]{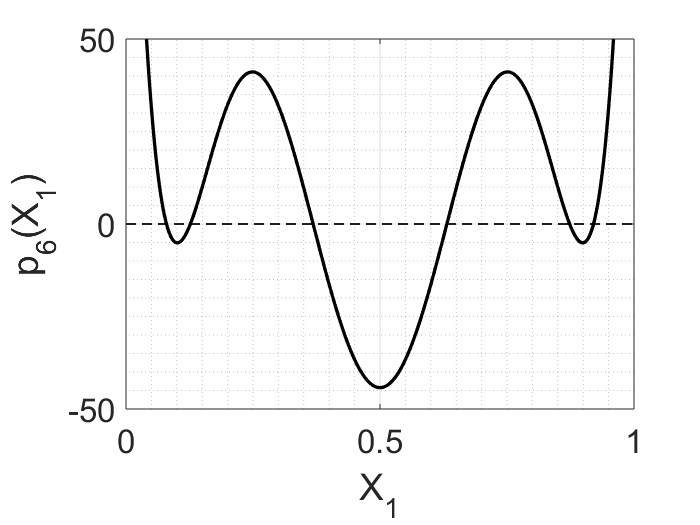}
		\caption{$\alpha=9,\ \beta=80$: $p_6(X_1)$ vs $X_1$}
		\label{case1aEg3p6}
    \end{subfigure}
	\caption{Three qualitatively different possibilities for Case I. In the left hand figures, nullcline \eqref{sub1a} is shown in blue and nullcline \eqref{sub2a} is shown in red. The pink region is the basin of attraction of the equilibrium $(X_1^e,Y_1^e)=(0,\frac 1{\alpha})$ and the blue region is the basin of attraction of $(X_1^e,Y_1^e)=(1,0)$. The white regions in figure~\ref{case1aEg3} correspond to the basin of attraction of two of the new equilibria in that figure. Stable nodes are denoted by $\CIRCLE$, unstable nodes by $\Circle$ and saddle points by $\bigotimes$. The right hand figures show $p_6(X_1)$ in each case.} 
	\label{fig:case1aEg}
\end{figure}

We show examples of three qualitatively different cases in figure~\ref{fig:case1aEg} for $(\alpha,\beta)=(9,16),(9,40),(9,80)$.
On the left hand side, we show the nullclines \eqref{sub1a}, \eqref{sub2a} in $(X_1,Y_1)$-space for each case. We colour the basins of attraction in our plots according to the stable equilibrium which it contains\footnote{In fact, the colour scheme is based on the \textit{index of dissimilarity} \cite{Massey1988} evaluated at the corresponding stable equilibrium state; see section~\ref{sec:disc}.}. On the right hand side of figure~\ref{fig:case1aEg}, we plot the corresponding $p_6(X_1)$. 

Figure~\ref{case1aEg1} shows the case when $\alpha=9,\ \beta=16$. Here $p_9(X_1)$ has only three real roots corresponding to $(X_1^e,Y_1^e)=(1,0), (0,\frac 1{\alpha}),\ (\frac 12,\frac 1{2\alpha})$. The basin of attraction of $(X_1^e,Y_1^e)=(0,\frac 1{\alpha})$ is shown in pink and the basin of attraction of $(X_1^e,Y_1^e)=(1,0)$ is shown in blue. The equilibrium  $(X_1^e,Y_1^e)=(\frac 12,\frac 1{2\alpha})$ has no basin of attraction. Figure~\ref{case1aEg1p6} shows that the corresponding $p_6(X_1)$ has no real zeros. 

In Figure~\ref{case1aEg2}, we take $\alpha=9,\ \beta=40$. Now nullclines \eqref{sub1a}, \eqref{sub2a} have five intersections, corresponding to five real roots of $p_9(X_1)$. Three of these roots correspond to $(X_1^e,Y_1^e)=(1,0), (0,\frac 1{\alpha}),\ (\frac 12,\frac 1{2\alpha})$, as before. So the two new roots correspond to zeros of $p_6(X_1)$. Figure~\ref{case1aEg2p6} shows that $p_6(X_1)$ has indeed developed two zeros. Neither of these two new equilibria has a basin of attraction\footnote{Note that the central equilibrium $(X_1^e,Y_1^e)=(\frac 12,\frac 1{2\alpha})$ has changed its character from a saddle in figure~\ref{case1aEg1} to an unstable node in figure~\ref{case1aEg2}. We discuss the nature of these equilibria in section~\ref{sec:caseIstab} below}.

In Figure~\ref{case1aEg3}, we take $\alpha=9,\ \beta=80$. Nullclines \eqref{sub1a}, \eqref{sub2a} have nine intersections, corresponding to nine real roots of $p_9(X_1)$ and $p_6(X_1)$ has six zeros, seen in figure~\ref{case1aEg3p6}. Two of these six roots correspond to integrated populations with a basin of attraction, shown in white in figure~\ref{case1aEg3}.

In this case, there is no example of $p_9(X_1)$ having seven roots, corresponding to $p_6(X_1)$ having four zeros.

From a societal perspective, figure~\ref{fig:case1aEg} paints a rather gloomy picture. Figures~\ref{case1aEg1} and \ref{case1aEg2} both imply that populations will be segregated at these values of $(\alpha,\beta)$, since both figures show only basins of attraction of segregated populations. The central (integrated) equilibrium  $(X_1^e,Y_1^e)=(\frac 12,\frac 1{2\alpha})$ never has a basin of attraction and so appears unstable. Stable integration seems possible only in figure~\ref{case1aEg3}, corresponding to a small, highly tolerant, minority. As in the single-neighbourhood case \cite{Haw2018}, {\it tipping points} of the system correspond to the boundaries of the basins of attraction. 

We will now establish exact criteria for the existence and stability of roots of $p_9(X_1)$ as $(\alpha,\beta)$ vary.

\subsubsection{Existence of equilibria of \eqref{eq:case1anondim}}
Since we always have roots $X_1^e=0, \frac 12,1$ of $p_9(X_1)$, we need only consider the existence of roots of $p_6(X_1)$. From the right hand side of figure~\ref{fig:case1aEg}, there will be at least two real roots of $p_6(X_1)$ when $$p_6(\frac{1}{2})<0.$$ Since $p_6(\frac{1}{2}) = \alpha^2-\frac{1}{2}\alpha\beta+\beta$, then $p_6(\frac{1}{2})<0$ implies that 
\begin{eqnarray}
\label{eq:p60p5}
\beta>\beta_c \equiv\frac{2\alpha^2}{\alpha-2}.
\end{eqnarray}

Since $\beta>0$, we must have $\alpha>2$. At $\beta = \beta_c$, the two nullclines \eqref{sub1a}, \eqref{sub2a} have a cubic tangency. Note the minimum value of $\beta_c=16$, which occurs when $\alpha=4$. The next step is to observe that if $X_1=\frac{1}{2}+\eta$ is a root of $p_6(X_1)=0$, then by symmetry, so too is $X_1=\frac{1}{2}-\eta$, where $\eta \in [0,\frac12]$. So we have two equations for $\eta$:
\begin{eqnarray}
\label{eq:p6eta}
\sum_{i=0}^6a_i(\frac{1}{2}\pm\eta)^i&=&0,
\end{eqnarray}
where $a_i, \ i=0 \ldots 6$ are given in \eqref{eq:ai}. We now expand both equations in \eqref{eq:p6eta}, and add to give
\begin{eqnarray}
\label{eq:eta}
A\eta^6+B\eta^4+C\eta^2+D&=&0
\end{eqnarray}
where
\begin{eqnarray}
\label{eq:ABCD}
A&=&8\alpha^2\beta,\\
B&=&4\alpha\beta(3-\alpha),\\
C&=&\frac12\beta(\alpha^2-6\alpha+12),\\
D&=&\alpha-\frac{\beta}{2}+\frac{\beta}{\alpha}.
\end{eqnarray}

So $\eta^2$ has either one or three real values, depending on the sign of the discriminant $\Delta \equiv B^2C^2-4AC^3-4B^3D-27A^2D^2+18ABCD$, because \eqref{eq:eta} is a cubic in $\eta^2$. Hence $p_6(X_1)$ will have either two real roots or six real roots; it can not have four real roots. Hence $p_9(X_1)=0$ can not have seven roots (the missing case from figure~\ref{fig:case1aEg}). Note that this is not in contradiction of Descartes rule of signs, which is a necessary condition only. Note that when $D=0$, we have $\beta=\beta_c$ and \eqref{eq:eta} has solution $\eta^2=0$. Hence $p_6(\frac{1}{2})=0$, which corresponds exactly to \eqref{eq:p60p5}.

A lengthy calculation shows that 
\begin{eqnarray}
\label{eq:disc}
\Delta&=&-4\alpha^5\beta^2\left[8\alpha^2\beta+\alpha(432-72\beta-\beta^2)+8\beta^2\right].
\end{eqnarray}
Hence $\Delta=0$ when $\beta=\beta_{\pm}$ where
\begin{eqnarray}
\label{eq:3real}
\beta_{\pm} \equiv \frac{4}{8-\alpha}\left[9\alpha-\alpha^2 \pm \sqrt{\alpha(\alpha-6)^3}\right].
\end{eqnarray}

So $\eta^2$ has three real values, and $\eta$ has six real values, for $\beta\in[\beta_-,\beta_+]$. At $\beta = \beta_{\pm}$, the two nullclines \eqref{sub1a}, \eqref{sub2a} have quadratic tangencies. We must have $\alpha \ge 6$ for real values of $\beta_{\pm}$.

\begin{figure}[h!]
	\centering
     \includegraphics[width=.7\textwidth]{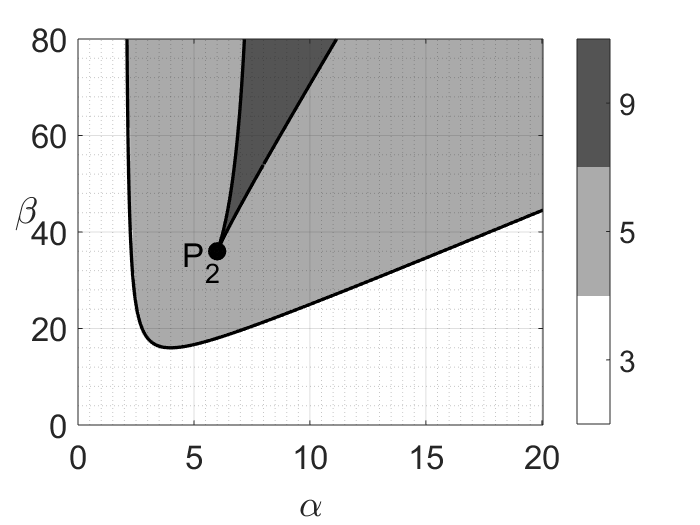}
	\caption{Number of equilibria of \eqref{eq:case1anondim}, corresponding to the roots of $p_9(X_1)$. Solid lines correspond to $\beta= \beta_c \equiv 2\alpha^2/(\alpha-2)$, from \eqref{eq:p60p5} and $\beta = \beta_{\pm} \equiv \frac{4}{8-\alpha}\left[9\alpha-\alpha^2 \pm \sqrt{\alpha(\alpha-6)^3}\right]$, from \eqref{eq:3real}. Integrated equilibria occur only inside the dark shaded region, for $\beta \in [\beta_-,\beta_+]$, when \eqref{eq:case1anondim} has nine equilibria. The polynomial $p_9(X_1)$ has five roots in the light shaded region and only three roots in the white shaded region. In both these cases, only segregated equilibria are possible. The apex $P_2$ of the dark shaded region is the point $(\alpha,\beta)=(6,36)$.}
	\label{pspace1}
\end{figure}

The existence of equilibria of \eqref{eq:case1anondim} is summarised in figure~\ref{pspace1}. Solid lines correspond to $\beta=\beta_c \equiv 2\alpha^2/(\alpha-2)$, from \eqref{eq:p60p5} and $\beta = \beta_{\pm} \equiv \frac{4}{8-\alpha}\left[9\alpha-\alpha^2 \pm \sqrt{\alpha(\alpha-6)^3}\right]$, from \eqref{eq:3real}. The apex of the dark shaded region is the point $P_2=(\alpha,\beta)=(6,36)$. Since integrated equilibria occur only inside this region, we can deduce that this version of the two-room problem requires a very tolerant (large $\beta$), small (large $\alpha$) minority to produce an integrated population (which may or may not be stable). 

The above analysis also allows us to obtain an exact expansion of $p_6(X_1)$ as a cubic in $(X_1-\frac{1}{2})$. Since $\eta = \pm (X_1-\frac{1}{2})$, we have from \eqref{eq:p6eta} and \eqref{eq:eta} that

\begin{eqnarray}
p_6(X_1) & = & \sum_{i=0}^6a_iX_1^i \\
& = & A(X_1-\frac{1}{2})^6 + B(X_1-\frac{1}{2})^4 + C (X_1-\frac{1}{2})^2 + D. \label{eq:p6root}
\end{eqnarray}
So
\begin{multline}
p_9(X_1)  =  X_1(1-X_1)(1-2X_1)\Biggl [ 8\alpha^2\beta\Bigl(X_1-\frac{1}{2}\Bigr)^6 + 4\alpha\beta(3-\alpha)\Bigl(X_1-\frac{1}{2}\Bigr)^4 \\
+ \frac12\beta(\alpha^2-6\alpha+12)\Bigl(X_1-\frac{1}{2}\Bigr)^2 + \Bigl(\alpha-\frac{\beta}{2}+\frac{\beta}{\alpha}\Bigr) \Biggr ] .
\label{eq:p6exp}
\end{multline}

Hence we can give exact expressions for all the roots $X_1=X_1^e$ of $p_9(X_1)=0$. We already have $X_1^e=0,\frac 12,1$. Now we can solve $p_6(X_1^e)=0$ for $(X_1^e-\frac{1}{2})^2$ from \eqref{eq:p6exp} using the standard formula for roots of a cubic, take the square root and obtain $X_1^e$ and then use \eqref{sub1a} to obtain the corresponding value of $Y_1^e$. These unwieldy expressions, not given here, can then be used to check the numerical results in figure~\ref{fig:case1aEg}. 

\subsubsection{Stability of equilibria of \eqref{eq:case1anondim}}
\label{sec:caseIstab}
It is clear from figure~\ref{fig:case1aEg} that not all equilbria of \eqref{eq:case1anondim} are stable, since they have no basin of attraction. In this section, we determine stability criteria for equilibria of \eqref{eq:case1anondim}. Let us write \eqref{eq:case1anondim} in the form 
\begin{eqnarray}
\label{eq:case1anondim1}
\frac{dX_1}{dt}&=& P(X_1,Y_1),\\
a_1\frac{dY_1}{dt}&=& Q(X_1,Y_1).\nonumber
\end{eqnarray}
where 
\begin{eqnarray}
\label{eq:PQ}
P(X_1,Y_1)&=&(1-X_1)[1-\alpha X_1 + 2\alpha X_1^2]-\alpha Y_1,\\
Q(X_1,Y_1)&=&(1-\alpha Y_1)[1-\beta Y_1+2\alpha\beta Y_1^2]-X_1.\nonumber
\end{eqnarray}

To establish stability criteria, we must calculate the eigenvalues of the Jacobian of  \eqref{eq:case1anondim1}, given by

\begin{eqnarray}
\label{eq:jac}
J(X_1,Y_1) \equiv
\left( 
\begin{array}{cc}
\frac{\partial P}{\partial X_1} & \frac{\partial P}{\partial Y_1} \\
\frac{\partial Q}{\partial X_1} & \frac{\partial Q}{\partial Y_1} 
\end{array} 
\right)
=
\left( 
\begin{array}{cc}
-(1+\alpha)+6\alpha X_1(1-X_1) & -\alpha \\
-1 & -(\alpha+\beta)+6\alpha \beta Y_1(1-\alpha Y_1) 
\end{array} 
\right),
\end{eqnarray}
evaluated at the various equilibria $(X_1,Y_1)=(X_1^e,Y_1^e)$.

For the equilibrium $(X_1^e,Y_1^e)=(1,0)$, the eigenvalues of $J(X_1,Y_1)$ are given by
\begin{eqnarray}
\label{eq:eigenvalues1}
\lambda_{\pm}&=&\frac{1}{2}
\left[-(1+2\alpha+\beta) \pm \sqrt{4\alpha+(1-\beta)^2}\right ].
\end{eqnarray}
It is straightforward to show that both $\lambda_{\pm}<0$ for all $\alpha>0,\beta>0$. Hence the equilibrium $(X_1^e,Y_1^e)=(1,0)$, corresponding to all of the $X$-population in neighbourhood $1$ and all of the $Y$-population in neighbourhood $2$, is a stable node, shown by a solid circle ($\CIRCLE$) in figure~\ref{fig:case1aEg}. Unless the system is modified in some way, this means that there will always be a non-empty set of initial conditions that will lead to this segregated outcome.

For the equilibrium $(X_1^e,Y_1^e)=(0,\frac{1}{\alpha})$, the eigenvalues are also given by \eqref{eq:eigenvalues1}. Similar considerations apply to this (stable) segregated outcome.

The equilibrium $(X_1^e,Y_1^e)=(\frac{1}{2},\frac{1}{2\alpha})$ has a more subtle behaviour. Its eigenvalues are given by
\begin{eqnarray}
\label{eq:eigenvalues2}
\lambda_{\pm}&=&\frac{1}{4}
\left[\beta-(\alpha+2) \pm \sqrt{(\beta+2)^2 + 9\alpha^2 + 4\alpha -6\alpha\beta}\right ].
\end{eqnarray}

We can show that $\lambda_+>0$ in \eqref{eq:eigenvalues2}. Hence the equilibrium is always unstable. In addition, $\lambda_-\gtrless0$ for $\beta\gtrless\beta_c$ where $\beta_c \equiv 2\alpha^2/(\alpha-2)$, from \eqref{eq:p60p5}. 
Hence $(X_1^e,Y_1^e)=(\frac{1}{2},\frac{1}{2\alpha})$ is a saddle for $\beta <\beta_c$, shown by a crossed circle ($\bigotimes$) in figure~\ref{fig:case1aEg}, and an unstable node for $\beta>\beta_c $, shown by an open circle ($\Circle$). At $\beta=\beta_c$, we have a supercritical pitchfork bifurcation, where the saddle at $(X_1^e,Y_1^e)=(\frac{1}{2},\frac{1}{2\alpha})$ becomes an unstable node and two saddles. This explains why no new stable equilibria are created at $\beta = \beta_c$. This bifurcation occurs precisely at the boundary between three and five equilibria shown in figure~\ref{pspace1}. 

The three equilibria considered so far are zeros of $p_9(X_1)$; they always exist for any values of $\alpha,\beta$. The remaining zeros of $p_9(X_1)$ come from $p_6(X_1)$. They can be found analytically, as explained above. But the expressions for the equilibria are unwieldy and the ensuing stability calculations are extremely lengthy. So we will simply summarise the results. 

\begin{figure}[h!]
	\centering
	\begin{subfigure}{.45\textwidth}
		\centering
		\includegraphics[width=\linewidth]{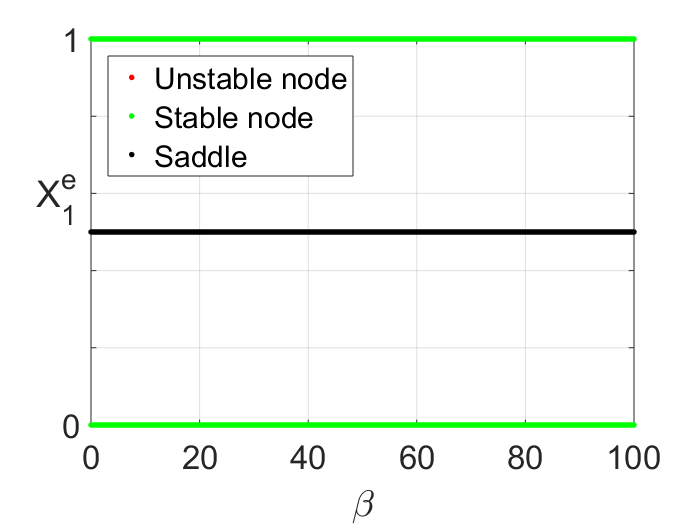}
		\caption{$\alpha=1.9$: $\beta_c,\, \beta_{\pm}$ undefined.}
		\label{bif1}
	\end{subfigure}
	\begin{subfigure}{.45\textwidth}
		\centering
		\includegraphics[width=\linewidth]{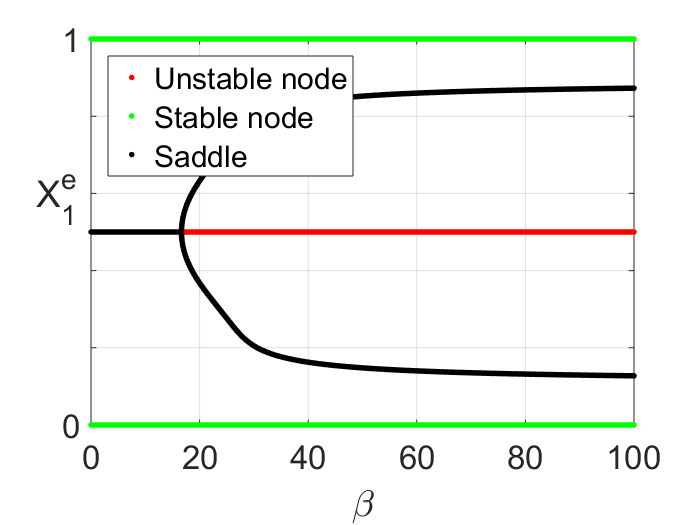}
		\caption{$\alpha=5,\,\beta_c=\frac{50}{3}\approx 16.67$: $\beta_{\pm}$ undefined.}
		\label{bif2}
	\end{subfigure} \\
	\begin{subfigure}{.45\textwidth} 
		\centering
		\includegraphics[width=\linewidth]{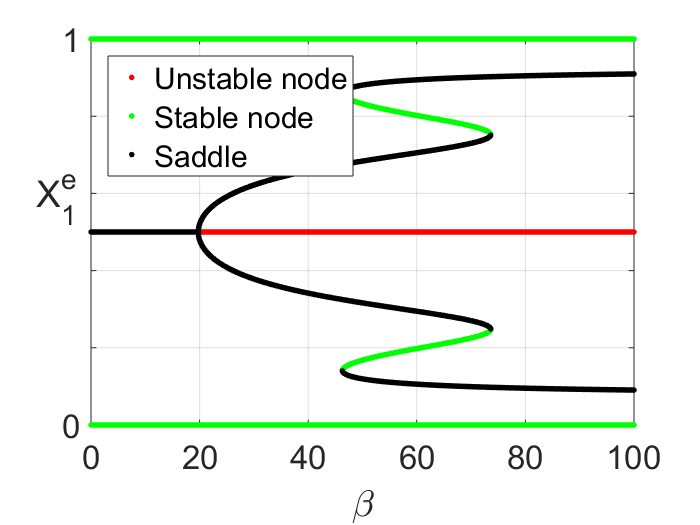}
		\caption{$\alpha=7.1$: $\beta_c \approx 19.77$ and $\beta _- \approx 46.29, \ \beta_+ \approx 73.62$.}
		\label{bif3}
	\end{subfigure}
	\begin{subfigure}{.45\textwidth} 
		\centering
		\includegraphics[width=\linewidth]{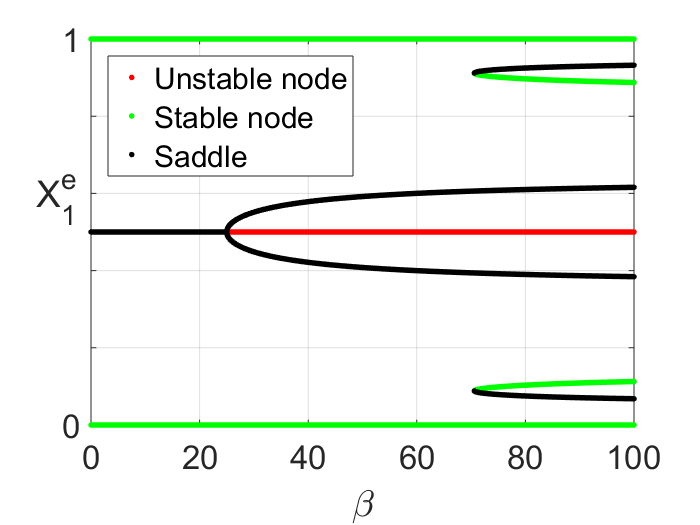}
		\caption{$\alpha=10$: $\beta_c =25$ and $\beta _- \approx 70.60$.}
		\label{bif4}
	\end{subfigure}
	\caption{Case I: equilibrium values $X_1^e$ as a function of $\beta$ for (a) $\alpha=1.9$, (b) $\alpha=5$, (c) $\alpha=7.1$ and (d) $\alpha=10$.}
	\label{bifdiags}
\end{figure}

In figure~\ref{bif1}, we set $\alpha=1.9$ and plot the resulting values of $X_1^e$ as a function of $\beta$ (the values of $Y_1^e$ are omitted, for convenience). Since $\alpha<2$, none of the quantities $\beta_c,\beta_{\pm}$ is defined and $p_9(X_1)$ only has three real zeros: $X_1^e=0,1$ (both stable nodes, shown in green) and $X_!^e=\frac{1}{2}$ (a saddle, shown in black).

In figure~\ref{bif2}, we take $\alpha=5$. Since $2<\alpha<6$, we have $\beta_c=\frac{50}{3}\approx 16.67$, with $\beta_{\pm}$ undefined. The saddle existing for $\beta < \beta_c$ becomes an unstable node (shown in red) and the new solutions for $\beta > \beta_c$ are saddles.

In figure~\ref{bif3}, we set $\alpha=7.1$. Hence $\beta_c \approx 19.77$ and $\beta _- \approx 46.29, \ \beta_+ \approx 73.62$. The pitchfork bifurcation at $\beta = \beta_c$ is followed by fold bifurcations at $\beta = \beta_{\pm}$. For $\beta \in [\beta_-,\beta_+]$, we see two stable nodes, corresponding to stable integration. These two solution branches have different basins of attraction (see figure~\ref{case1aEg3}).

Finally in figure~\ref{bif4}, $\alpha=10$ and now the pitchfork bifurcation is at $\beta=\beta_c =25$.  We can always find an integrated population whenever $\beta > \beta_- \approx 70.60$.

So to get a stable integration in two neighbourhoods with the same linear tolerance schedules, we need to select parameters $(\alpha, \beta)$ to lie in the dark shaded region of figure~\ref{pspace1}, corresponding to a small, highly tolerant, minority. For $6<\alpha<8$, we can only find an integrated solution when $\beta_-<\beta<\beta_+$.  For $\alpha>8$, we can always find an integrated population whenever $\beta>\beta_-$, provided we start with the right initial conditions. 

\subsection{Case II, III and IV}
In the previous section, the linear tolerance schedules \eqref{eq:toli} were the same in both neighbourhoods ($\gamma = 1$, $\beta_1 = \beta_2$) for both majority $X$- and minority $Y$-populations. Let us now consider the case in which the linear tolerance schedules are different. The types of people remain the same, but the two neighbourhoods induce a different tolerance in each population (this may be due to other factors such as urban environment, educational provision, etc). We revert to the original dynamical system \eqref{eq:case1nondim} and nullclines \eqref{sub1} and \eqref{sub2}. We distinguish three different cases:
\begin{itemize}
\item[Case II; $\gamma = 1$, $\beta_1 \ne \beta_2$:]  the majority population have the same tolerance in both neighbourhoods; the minority have different tolerances in both neighbourhoods.
\item[Case III; $\gamma \ne 1$, $\beta_1 = \beta_2$:]  the majority population have different tolerances in both neighbourhoods; the minority have the same tolerance in both neighbourhoods.
\item[Case IV; $\gamma \ne 1$, $\beta_1 \ne \beta_2$:]  both populations have different tolerances in both neighbourhoods.
\end{itemize}

Substituting \eqref{sub1} into \eqref{sub2}, we obtain a ninth order polynomial $\Tilde{p}_9(X_1)$ of possible equilibria. We know that $X_1^e=0,1$, by inspection of \eqref{sub1}, \eqref{sub2}. But $X_1^e = \frac{1}{2}$ is no longer a guaranteed equilibrium. Hence we  write $$\Tilde{p}_9(X_1)=X_1(1-X_1)\Tilde{p}_7(X_1),$$ where 
\begin{equation}
\label{eq:p7}
    \Tilde{p}_7(X_1) \equiv \sum_{i=0}^7b_iX_1^i.
\end{equation}
The real coefficients $b_i$ are given by
\begin{eqnarray}
b_0&=&-(\alpha+\beta_1)\gamma-\frac{\beta_1}{\alpha}\\
b_1&=&(1+\frac 1{\alpha})\beta_1+\alpha(1+\gamma)+\frac{\beta_2}{\alpha}+(5\beta_1+2\beta_2)\gamma+\alpha(2\beta_1+\beta_2)\gamma^2\nonumber\\
b_2&=&-2(2\beta_1+\beta_2)-[(7+4\alpha)\beta_1+(5+2\alpha)\beta_2]\gamma-3\alpha(3\beta_1+2\beta_2)\gamma^2-\alpha^2(\beta_1+\beta_2)\gamma^3\nonumber\\
b_3&=&(3+2\alpha)\beta_1+(3+\alpha)\beta_2+[(3+14\alpha)\beta_1+(3+10\alpha)\beta_2]\gamma+[3\alpha(5+\alpha)\beta_1+3\alpha(4+\alpha)\beta_2]\gamma^2+5\alpha^2(\beta_1+\beta_2)\gamma^3\nonumber\\
b_4&=&-\alpha(5\beta_1+4\beta_2)-\alpha[(16+3\alpha)\beta_1+(14+3\alpha)\beta_2]\gamma-\alpha[(11+12\alpha)\beta_1+(10+12\alpha)\beta_2]\gamma^2-10\alpha^2(\beta_1+\beta_2)\gamma^3\nonumber\\
b_5&=&\alpha(\beta_1+\beta_2)[3+\alpha+3(2+3\alpha)\gamma+3(1+6\alpha)\gamma^2+10\alpha\gamma^3]\nonumber\\
b_6&=&-\alpha^2(\beta_1+\beta_2)(2+9\gamma+12\gamma^2+5\gamma^3)\nonumber\\
b_7&=&\alpha^2(\beta_1+\beta_2)(1+\gamma)^3\nonumber
\end{eqnarray}

Note that $\Tilde{p}_7(X_1)$ evaluated at $\gamma=1, \ \beta_1=\beta_2$ can be shown to equal $(1-2X_1)p_6(X_1)$, as expected.

\subsubsection{Equilibria}
Equilibria of the governing equations \eqref{eq:case1nondim}, segregated or integrated, correspond to real zeros of $\Tilde{p}_9(X_1)$. We know by inspection that $X^e=0,1$ are equilibria. But owing to the lack of symmetry between the two neighbourhoods, we can say very little analytically about any other equilibria, corresponding to real zeros of $\Tilde{p}_7(X_1)$ in \eqref{eq:p7}. 
  
The signs of the coefficients $b_i$ of $\Tilde{p}_7(X_1)$ alternate  for allowed values of the parameters. So Descartes' rule of signs tells us that $\Tilde{p}_7(X_1)$ has 1,3,5 or 7 real roots for $X_1>0$ and no real roots for $X_1<0$. Hence $\Tilde{p}_9(X_1)$ has 3,5,7 or 9 real roots for $X_1>0$ and no real roots for $X_1<0$, as in case I.

We can also show that there has to be at least one real root of $\Tilde{p}_7(X_1)$ for $X_1 \in (0,1)$, for any allowed parameter values. Simple calculation shows that $\Tilde{p}_7(0) = -(\alpha+\beta_1)\gamma-\frac{\beta_1}{\alpha} <0$ and $\Tilde{p}_7(1) = (\alpha+\beta_2)+\frac{\beta_2}{\alpha} >0$. Since $\Tilde{p}_7(X_1)$ is continuous in $X_1$, the Intermediate Value Theorem tells us that there always has to be at least one zero of $\Tilde{p}_7(X_1)$ between $X_1=0$ and $X_1=1$. Since neither $\Tilde{p}_7(1)$ nor $\Tilde{p}_7(0)$ is identically zero, the root must lie strictly between $X_1=0$ and $X_1=1$.

Governing equations \eqref{eq:case1nondim} can be written in the form
\begin{eqnarray}
\label{eq:case1nondimgen}
\frac{dX_1}{dt} & = & K(X_1,Y_1),\\
a_1\frac{dY_1}{dt} & = & L(X_1,Y_1). \nonumber
\end{eqnarray}
where
\begin{eqnarray}
\label{eq:KL}
K(X_1,Y_1) & = & (1-X_1)[1-\alpha\gamma X_1 + \alpha(1+\gamma)X_1^2]-\alpha Y_1,\\
L(X_1,Y_1) & = & (1-\alpha Y_1)[1-\beta_2Y_1+\alpha(\beta_1+\beta_2)Y_1^2]-X_1. \nonumber
\end{eqnarray}

The Jacobian of \eqref{eq:case1nondimgen} is given by

\begin{multline}
\label{eq:jacgen}
J(X_1,Y_1) \equiv
\left( 
\begin{array}{cc}
\frac{\partial K}{\partial X_1} & \frac{\partial K}{\partial Y_1} \\
\frac{\partial L}{\partial X_1} & \frac{\partial L}{\partial Y_1} 
\end{array} 
\right)\\
=
\left( 
\begin{array}{cc}
-(1+\alpha\gamma)+2\alpha (1+2\gamma)X_1-3\alpha(1+\gamma)X_1^2 & -\alpha \\
-1 & -(\alpha+\beta_2)+2\alpha (\beta_1+2\beta_2) Y_1-3\alpha^2(\beta_1+\beta_2) Y_1^2)
\end{array} 
\right).
\end{multline}

For the equilibrium $(X_1^e,Y_1^e)=(1,0)$, the eigenvalues of $J(X_1,Y_1)$ are given by
\begin{eqnarray}
\label{eq:eigenvaluesagen}
\lambda_{\pm}&=&\frac{1}{2}
\left[-(1+2\alpha+\beta_2) \pm \sqrt{4\alpha+(1-\beta_2)^2}\right ] .
\end{eqnarray}
Both $\lambda_{\pm}<0$ and are independent of $\beta_1$. Hence $(X_1^e,Y_1^e)=(1,0)$ is always a stable node. When $\gamma=1,\,\beta_1=\beta_2=\beta$, \eqref{eq:eigenvaluesagen} reduces to \eqref{eq:eigenvalues1}.

For the equilibrium $(X_1^e,Y_1^e)=(0,\frac{1}{\alpha})$, the eigenvalues of $J(X_1,Y_1)$ are given by
\begin{eqnarray}
\label{eq:eigenvaluesbgen}
\lambda_{\pm}&=&\frac{1}{2}
\left[-(1+\beta_1)-\alpha(1+\gamma)
\pm \sqrt{\alpha^2(\gamma-1)^2+(\beta_1-1)^2+2\alpha[(\gamma+1)+\beta_1(1-\gamma)]}\right ].
\end{eqnarray}
These eigenvalues are both negative for all allowed values of $\alpha, \beta$ and are independent of $\beta_2$. Hence $(X_1^e,Y_1^e)=(0,\frac{1}{\alpha})$ is always a stable node. Equation \eqref{eq:eigenvaluesbgen} reduce to \eqref{eq:eigenvalues1} when $\gamma=1,\,\beta_1=\beta_2=\beta$. 

Our considerations above show that there is always at least one other equilibrium of \eqref{eq:case1nondimgen} in addition to $(X_1^e,Y_1^e)=(1,0),\, (0,\frac{1}{\alpha})$. In the case when there is exactly one additional equilibrium, the fact that both  $(X_1^e,Y_1^e)=(1,0),\, (0,\frac{1}{\alpha})$ are always stable nodes means that this third equilibrium must be a saddle. 

In general, owing to the lack of symmetry, further equilibria must be created by fold bifurcations. These happen when turning points  of $\Tilde{p}_7(X_1)$ (locally quadratic maxima or minima) cross the $X_1$ axis. We do not expect pitchfork bifurcations in case II, III or IV, which occur in systems with an inversion or reflection symmetry \cite{Guckenheimer97}, for example case I.

Fold bifurcations produce either a stable node (a desirable outcome because $X_1^e \ne 0$ and $Y_1^e \ne 0$) and a new saddle or an unstable node and a new saddle. We then face the following possibilities: (i) further equilibria (stable or unstable) can be produced by additional new fold bifurcations,  (ii) the original saddle can disappear in a fold bifurcation with the newly created node or (iii) the original fold bifurcation can be reversed. 

\begin{figure}[h!]
	\centering
	\begin{subfigure}{.45\textwidth}
		\centering
		\includegraphics[width=\linewidth]{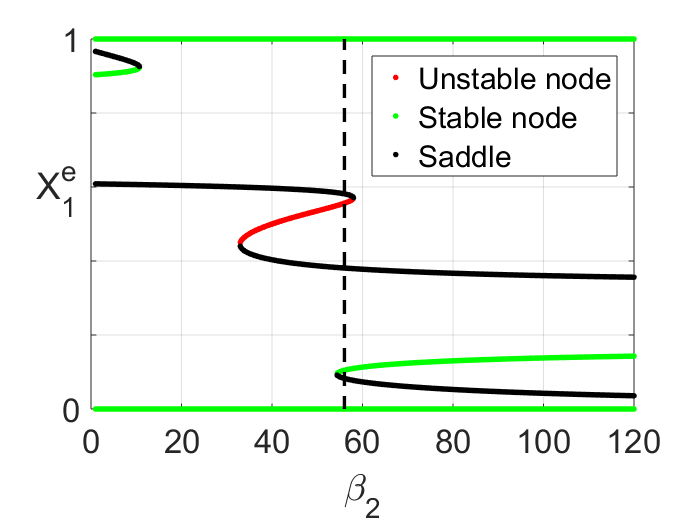}
		\caption{Case II: $\gamma=1,\,\beta_1 \ne \beta_2$; $\alpha=9,\, \beta_1=40$.}
		\label{caseIIbifdiag}
	\end{subfigure}
	\begin{subfigure}{.45\textwidth}
		\centering
		\includegraphics[width=\linewidth]{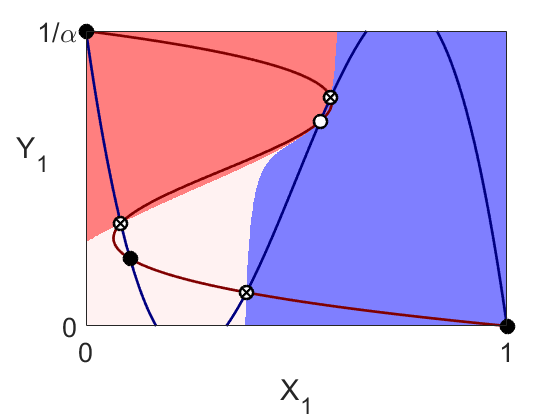}
		\caption{Case II with $\alpha=9,\, \beta_1=40,\,\beta_2=56$.}
		\label{caseIIphase}
	\end{subfigure}\\
	\begin{subfigure}{.45\textwidth}
		\centering
		\includegraphics[width=\linewidth]{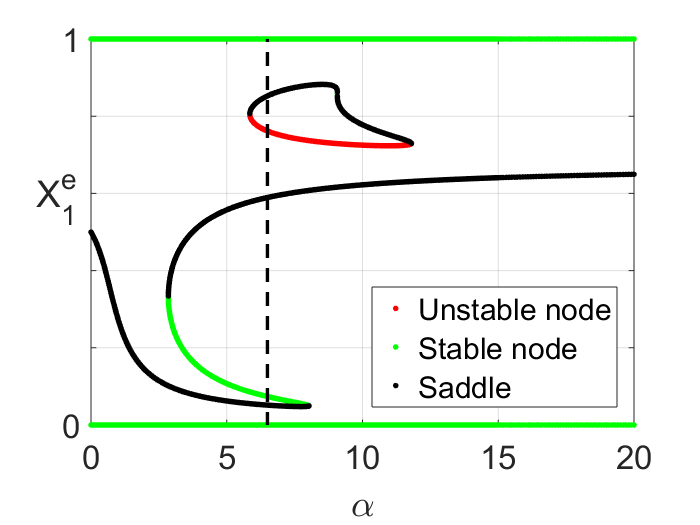}
		\caption{Case III: $\gamma \ne 1,\ \beta_1=\beta_2$; $\gamma=2,\ \beta_1=\beta_2=60$.}
		\label{caseIIIbifdiag}
	\end{subfigure}
	\begin{subfigure}{.45\textwidth}
		\centering
		\includegraphics[width=\linewidth]{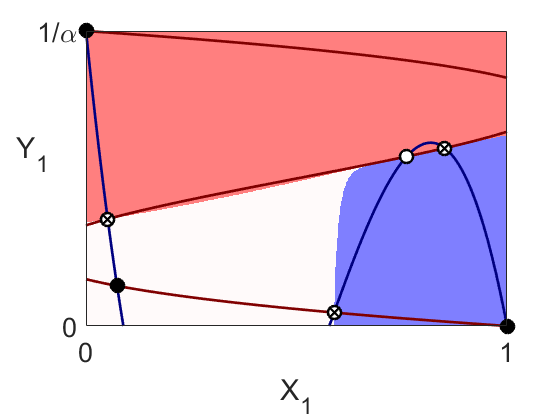}
		\caption{Case III with $\gamma=2,\ \beta_1=\beta_2=60,\,\alpha=6.5$.}
		\label{caseIIIphase}
	\end{subfigure}\\
	\begin{subfigure}{.45\textwidth} 
		\centering
		\includegraphics[width=\linewidth]{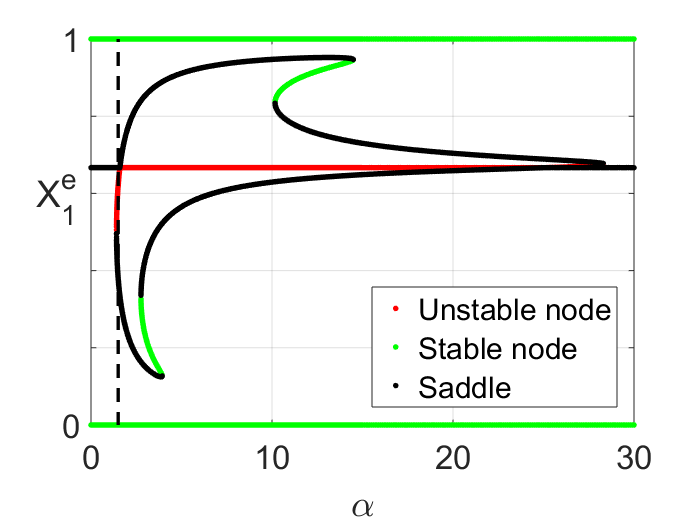}
		\caption{Case IV: $\gamma \ne 1,\ \beta_1 \ne \beta_2$; $\gamma=2,\ \beta_1=80,\ \beta_2=40$.}
		\label{caseIVbifdiag}
    \end{subfigure}
	\begin{subfigure}{.45\textwidth}
		\centering
		\includegraphics[width=\linewidth]{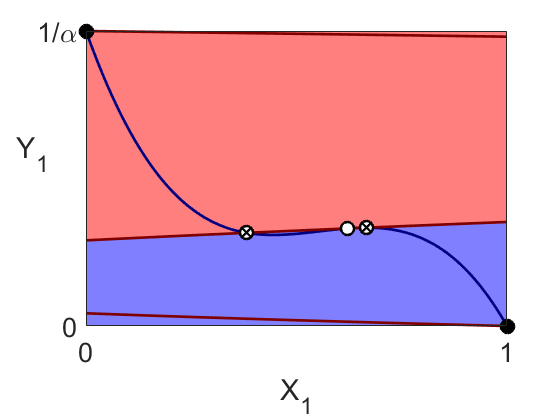}
		\caption{Case IV with $\gamma=2,\ \beta_1=80,\ \beta_2=40,\,\alpha=1.5$.}
		\label{caseIVphase}
	\end{subfigure}
	\caption{Bifurcation diagrams and example phase portraits for case II, III and IV. Stable equilibria are denoted by $\CIRCLE$, unstable nodes by $\Circle$ and saddle points by $\bigotimes$.}
	\label{fig:case1bEg}
\end{figure}

Finally in this section, we show some bifurcation diagrams to illustrate the wealth of possible behaviour that can occur. For case II, $\gamma = 1$, $\beta_1 \ne \beta_2$, we take $\alpha=9,\ \beta_1=40$ and vary $\beta_2$, shown in figure~\ref{caseIIbifdiag}. The phase space diagram for $\beta_2=56$ is shown in figure~\ref{caseIIphase}. For case III, $\gamma \ne 1$, $\beta_1 = \beta_2$, we take $\gamma=2,\ \beta_1=\beta_2=60$ and vary $\alpha$, shown in figure~\ref{caseIIIbifdiag}. The phase space diagram for $\alpha=6.5$ is shown in figure~\ref{caseIIIphase}. Finally for case IV, $\gamma \ne 1$, $\beta_1 \ne \beta_2$, we take $\gamma=2,\ \beta_1=80,\ \beta_2=40$ and vary $\alpha$, as shown in figure~\ref{caseIVbifdiag}. Note the presence of a transcritical bifurcation near $\alpha=1.5$, which arises due to local symmetry. The phase space diagram for $\alpha=1.5$ is shown in figure~\ref{caseIVphase}. 


\section{Limiting numbers}
\label{sec:limit}
For the case of a single neighbourhood, when parameters $\alpha, \, \beta$ are such that only segregation is possible, Schelling \cite{Schelling1969, Schelling1971} proposed that an integrated population could be obtained by limiting numbers of one or both populations. In \cite{Haw2018}, we gave  exact conditions under which this could occur. We also analysed the stability of the resulting equilibria and showed that the removal of the most intolerant individuals can lead to integration for most values of $\alpha, \, \beta$. For some parameter values, it is possible to create up to seven new equilibria. But there are some values of $\alpha, \, \beta$ where limitation of the population  can not produce integration. 

In this section, we consider how limiting numbers might affect the population mixture for two neighbourhoods in case I (cases II, III and IV can be treated similarly). The picture is considerably more complicated than the one neighbourhood case. There are at least two ways to limit the population when there is more than one neighbourhood. We can restrict the overall number of one population.
So for example, we could take $X_1+X_2=u$ for $u \in (0,1)$. This is equivalent to the solution proposed by Schelling \cite{Schelling1969, Schelling1971} in the case of one neighbourhood. Note that this is not the same as a simple rescaling the $X$-population, since the least tolerant member of the $X$-population can now abide a $\frac{Y}{X}$ ratio of $(1-u) \ne 0$.

Another way to restrict population when there is more than one neighbourhood is to impose a limit in one neighbourhood only, for example $X_1=u<1$, but keep the overall population unchanged. Consequently in this case $X_2 \in [1-u,1]$, since $X_1+X_2=1$ from \eqref{eq:x1x2}. This way of limiting population is different to that treated by Schelling \cite{Schelling1969, Schelling1971} and \cite{Haw2018}, so we will analyse it here.

\begin{figure}[h!]
\centering
		\centering
		\includegraphics[width=.7\linewidth]{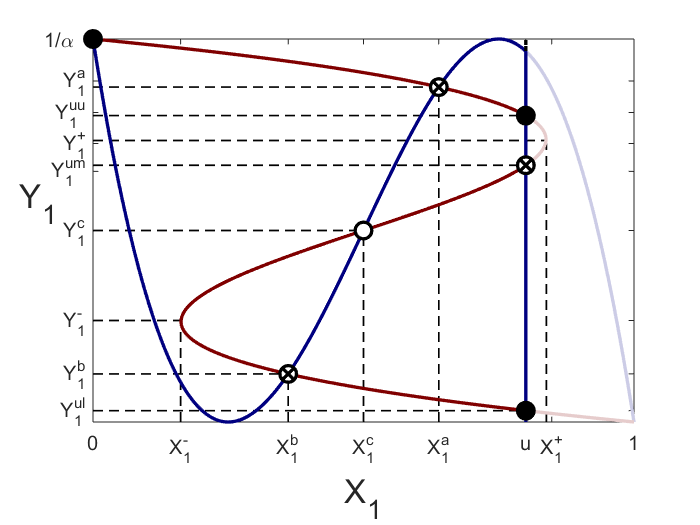}
        \caption{Limiting the $X_1$ population: $X_1=u$: $\beta \in [2\alpha,8\alpha]$ , $Y_1^a>Y_1^+$.}
        \label{fig:photolimitnumbersbgt2a}
\end{figure}

Let us begin by restricting the $X_1$-population (and hence the $X_2$-population), as illustrated in figure~\ref{fig:photolimitnumbersbgt2a}.
The $Y$-population is not restricted. Integration will correspond to intersections of the  line $X_1=u$ with the cubic nullcline \eqref{sub2a} given by $X_1=(1-\alpha Y_1)[1-\beta Y_1+2\alpha\beta Y_1^2]$. In figure~\ref{fig:photolimitnumbersbgt2a}, we show the case $\beta \in [2\alpha,8\alpha]$, when nullcline \eqref{sub2a} has a middle branch, which lies completely within the feasible $(X_1,Y_1)$ phase plane. In the absence of any population restriction, there will be three integrated equilibria, given by $(X_1,Y_1)= (X_1^e,Y_1^e)=(X_1^a,Y_1^a),\,(X_1^c,Y_1^c),\,(X_1^b,Y_1^b)$. 

The turning points $(X_1^{\pm},Y_1^{\pm})$ of nullcline \eqref{sub2a} are important. If the $X$-population is restricted at $X_1=u$, we exclude the most intolerant people. But for $u \in [X_1^+,1)$, we do not gain any extra equilibrium. Instead the segregated equilbrium at $(X_1,Y_1)=(0,1)$ becomes the (slightly) integrated equilbrium $(X_1,Y_1)=(u,Y_1^{ul})$. 

Figure~\ref{fig:photolimitnumbersbgt2a} shows the case when the upper unrestricted equilibrium $Y_1^a>Y_1^+$ with $u \in [X_1^a,X_1^+]$. There are three new equilbria, denoted by $(X_1,Y_1)=(u,Y_1^{ul}),\, (u,Y_1^{um}),\,(u,Y_1^{uu})$ on the lower, middle and upper branches, respectively, where $Y_1=Y_1^{u(l,m,u)}$ are the real roots of the cubic equation 
\begin{eqnarray}
\label{eq:cubiclimit}
u&=&(1-\alpha Y_1)[1-\beta Y_1+2\alpha\beta Y_1^2]
\end{eqnarray} 
when the discriminant of \eqref{eq:cubiclimit} is positive.
Note that if $Y_1^a<Y_1^+$, we can only get two new equilibria if $u \in [X_1^a,X_1^+]$ (not shown).

So to proceed we must first find out when the turning points $(X_1^{\pm},Y_1^{\pm})$ of nullcline \eqref{sub2a} exist. Then since the case $Y_1^a=Y_1^+$ (and hence by symmetry\footnote{In cases II, III and IV, there will be two different intersections, due to the lack of symmetry.} $Y_1^b=Y_1^-$) separates different types of behaviour, we must investigate when the two nullclines \eqref{sub1a}, \eqref{sub2a} intersect there. 

Turning points $(X_1^{\pm},Y_1^{\pm})$ of nullcline \eqref{sub2a} exist when it has a vertical tangent. From \eqref{sub2a}, we find

\begin{eqnarray}
\frac{dY_1}{dX_1} & = & \frac{1}{[-(\alpha + \beta)+6\alpha\beta Y_1-6\alpha^2\beta Y_1^2]}.
\end{eqnarray}

It is straightforward to show that, when $\beta > 2\alpha$, the nullcline \eqref{sub2a} has vertical tangents at

\begin{eqnarray}
 (X_1^{\pm},Y_1^{\pm}) & = & \left (\frac12 \pm \frac{1}{18\alpha\beta}\sqrt{3\beta(\beta-2\alpha)^3},\frac{1}{2\alpha} \pm \frac{1}{6\alpha\beta}\sqrt{3\beta(\beta-2\alpha)} \right ).
\end{eqnarray}
When $\beta \in [2\alpha,8\alpha]$, these vertical tangents lie within the feasible $(X_1,Y_1)$-plane. When $\beta = 2\alpha$, the two nullclines \eqref{sub1a}, \eqref{sub2a}, have a cubic tangency at the central equilibrium $(X_1,Y_1)=(X_1^c,Y_1^c)=(X_1^+,Y_1^+)=(X_1^-,Y_1^-)=(\frac12,\frac{1}{2\alpha})$.

 
 

Let us now investigate ways in which equilibria can occur on the middle branch of nullcline \eqref{sub2a}, when $\beta \in [2\alpha,8\alpha]$. Our aim is to find a curve $\Gamma_u(\alpha,\beta)=0$ that separates regions where three integrated equilibria are possible from regions where two integrated equilibria are possible. Points on this curve must satisfy:
\begin{equation}
\label{eq:ulimitingX1}
X_1^{\pm} \equiv \frac12 \pm \frac{1}{18\alpha\beta}\sqrt{3\beta(\beta-2\alpha)^3}=u,
\end{equation}
\begin{equation}
\label{eq:ulimitingY1}
\alpha Y_1^{\pm} \equiv \frac12 \pm \frac{1}{6\beta}\sqrt{3\beta(\beta-2\alpha)} = (1-u)(1-\alpha u+2\alpha u^2).
\end{equation}

Equation \eqref{eq:ulimitingX1} is the statement that the lines $X_1=u$ and $X_2=1-u$ intersect, due to symmetry, the vertical tangents of nullcline \eqref{sub2a}. Equation \eqref{eq:ulimitingY1} is the statement that points $(u,Y_1^{\pm})$ lie on nullcline \eqref{sub2a}.

Substituting \eqref{eq:ulimitingX1} into \eqref{eq:ulimitingY1}, it can be shown that
\begin{eqnarray}
\label{eq:gammau}
\Gamma_u(\alpha,\beta) & \equiv & \beta^3-9\alpha\beta^2+6\alpha(\alpha+9)\beta+16\alpha^3.
\end{eqnarray}

\begin{figure}[h!]
\centering
\includegraphics[width=0.7\linewidth]{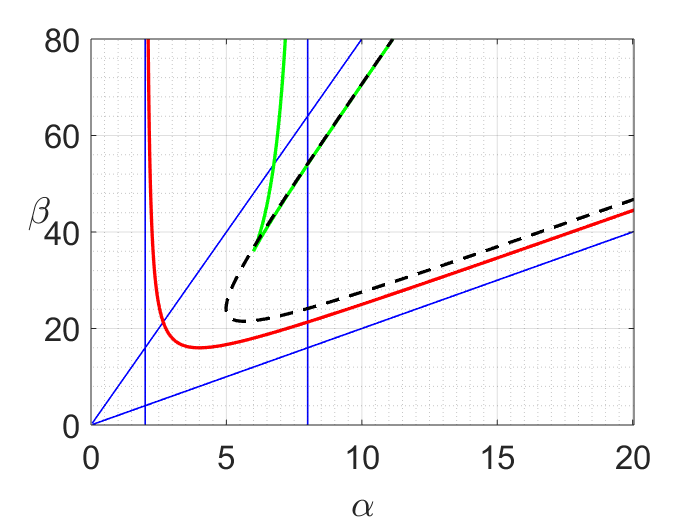}
\caption{Limiting the $X_1$-population. $\Gamma_u(\alpha,\beta)=0$  (equation \eqref{eq:gammau}, shown by black dashed line); $\beta=\beta_c$ (equation \eqref{eq:p60p5}, shown in red); $\beta=\beta_{\pm}$ (equation \eqref{eq:3real}, shown in green), together with the lines $\alpha=2,\, \alpha=8$ and $\beta=2\alpha,\,\beta=8\alpha$.}
\label{fig:limitnumbers}
\end{figure}

The curve $\Gamma_u(\alpha,\beta)=0$ is shown by the black dashed line in figure~\ref{fig:limitnumbers}. Note that it is asymptotic to $\beta=\beta_c$ and $\beta=\beta_-$, as $\alpha \to \infty$.
Also shown in the same figure are $\beta=\beta_c$ (equation \eqref{eq:p60p5}, shown in red); $\beta=\beta_{\pm}$ (equation \eqref{eq:3real}, shown in green), together with the lines $\alpha=2,\, \alpha=8$ and $\beta=2\alpha,\,\beta=8\alpha$ (shown in blue). We can see that the $(\alpha,\beta)$ parameter space is then divided up into 19 regions (some of which are extremely small). In each of these regions, the effect of restricting the $X_1$-population is slightly different. We shall discuss below the behaviour in two of these regions.

Let us now consider the case when we limit the $Y_1$-population (and hence the $Y_2$-population), but not the $X$-population. New integrated equilibria will correspond to intersections of the line $Y_1=v$ with the cubic nullcline \eqref{sub1a}, given by solutions $Y_1=Y_1^v$ of $\alpha Y_1=(1-X_1)[1-\alpha X_1+2\alpha X_1^2]$ (not shown). In a similar manner to above, it can be shown that turning points\footnote{We retain the same notation for these turning points. No confusion should arise.} $(X_1^{\pm},Y_1^{\pm})$ are given by
\begin{eqnarray}
 (X_1^{\pm},Y_1^{\pm}) & = & \left (\frac12 \pm \frac{1}{6\alpha}\sqrt{3\alpha(\alpha-2)},\frac{1}{2\alpha} \pm \frac{1}{18\alpha^2}\sqrt{3\alpha(\alpha-2)^3} \right ),
\end{eqnarray}
which exist whenever $\alpha>2$. When $\alpha=2$, the two nullclines \eqref{sub1a}, \eqref{sub2a}, have a cubic tangency at the central equilibrium $(X_1,Y_1)=(X_1^c,Y_1^c)=(X_1^+,Y_1^+)=(X_1^-,Y_1^-)=(\frac12,\frac{1}{2\alpha})$.
It can be shown that the middle branch of nullcline \eqref{sub2a} exists wholly with the permitted phase plane region when $\alpha \in [2,8]$. 

Our aim is to find a curve $\Gamma_v(\alpha,\beta)=0$ in the $(\alpha,\beta)$ plane that separates regions where three integrated equilibria are possible from regions where two integrated equilibria are possible. Points on this
curve must satisfy:
\begin{equation}
\label{eq:vlimitingY1}
Y_1^{\pm} \equiv \frac12 \pm \frac{1}{18\alpha^2}\sqrt{3\alpha(\alpha-2)^3}=v,
\end{equation}
\begin{equation}
\label{eq:vlimitingX1}
X_1^{\pm} \equiv \frac12 \pm \frac{1}{6\alpha}\sqrt{3\alpha(\alpha-2)} = (1-\alpha v)(1-\beta v+2\alpha \beta v^2).
\end{equation}
Equation \eqref{eq:vlimitingY1} is the statement that the line $Y_1=v$ intersects both horizontal tangents of nullcline \eqref{sub1a}. Equation \eqref{eq:vlimitingX1} is the statement that points $(X_1^{\pm},v)$ lie on nullcline \eqref{sub1a}.

It can be shown that
\begin{eqnarray}
\label{eq:gammav}
\Gamma_v(\alpha,\beta) & \equiv & \beta - \frac{54\alpha^2}{(8-\alpha)(\alpha-2)(\alpha+1)}.
\end{eqnarray}
We do not show the curve $\Gamma_v(\alpha,\beta)=0$ in  figure~\ref{fig:limitnumbers}. It is asymptotic to both the lines $\alpha=2,\,8$.

The effect of restricting the $Y_1$-population can be seen in $(\alpha,\beta)$ parameter space, by replacing the curve $\Gamma_u(\alpha,\beta)=0$ by $\Gamma_v(\alpha,\beta)=0$ in figure~\ref{fig:limitnumbers}. The effect of restricting both the $X_1$- and $Y_1$-populations can be seen by adding the curve $\Gamma_v(\alpha,\beta)=0$ to figure~\ref{fig:limitnumbers}. 

Both $\Gamma_{u,v}(\alpha,\beta)=0$ have a minimum at the same value of $\beta$. The minimum of $\Gamma_u$ occurs at the point $(\alpha_u,\beta) \approx (3.798, 21.488)$, where nullcline \eqref{sub2} passes through both the maximum and minimum of nullcline \eqref{sub1}. Similarly, the minimum of $\Gamma_v$ occurs at the point $(\alpha_v,\beta) \approx (5.658, 21.488)$, where nullcline \eqref{sub1} passes through both the maximum and minimum of nullcline \eqref{sub2}. 

So far in this section, we have only considered the existence of new equilibria when a population is limited. Let us now consider the stability of these new solutions. When the $X_1$-population is restricted to $X_1=u$, the dynamics on this line is governed by the second equation in each of \eqref{eq:case1nondimgen} and \eqref{eq:KL}:
\begin{equation}
\label{eq:limitXstab}
a_1\frac{dY_1}{dt}  =  L(u,Y_1)  =  (1-\alpha Y_1)[1-\beta Y_1+2\alpha\beta Y_1^2]-u.
\end{equation}
The equilibrium $Y_1=Y_1^u$ of \eqref{eq:limitXstab} is the solution of  \eqref{eq:cubiclimit}. There are either one or three real values of $Y_1^u$, depending on whether the line $X_1=u$ crosses the nullcline once or three times. Stability is governed by the eigenvalue $$\lambda_u=-(\alpha+\beta)+6\alpha\beta Y_1^u-6\alpha^2\beta(Y_1^u)^2.$$ Analytical progress can be made in finding $\lambda_u$, but in general it has to be evaluated numerically. System \eqref{eq:limitXstab} can never undergo a Hopf bifurcation to a periodic solution since it is only one-dimensional.
Similar conclusions apply when the $Y_1$-population is restricted to $Y_1=v$. The dynamics on this line are governed by the first equation in each of \eqref{eq:case1nondimgen} and \eqref{eq:KL}:
\begin{equation}
\frac{dX_1}{dt}  =  K(X_1,v)= (1-X_1)[1-\alpha X_1 + 2\alpha X_1^2]-\alpha v,\\
\end{equation}
and the stability of the equilibrium $X_1=X_1^v$ is governed by the eigenvalue $$\lambda_v-(1+\alpha)+6\alpha X_1^v-6\alpha (X_1^v)^2,$$ which can be evaluated numerically. 

\begin{figure}[h!]
\centering
\begin{subfigure}{.45\textwidth}
		\centering
		\includegraphics[width=\linewidth]{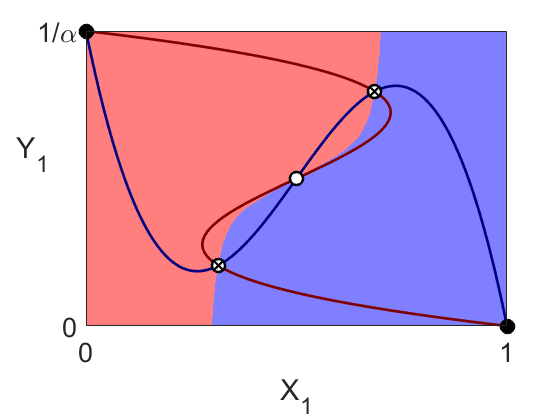}
        \caption{$u=1$}
        \label{fig:a6b30u1p0}
\end{subfigure}
\begin{subfigure}{.45\textwidth}
		\centering
    \includegraphics[width=\linewidth]{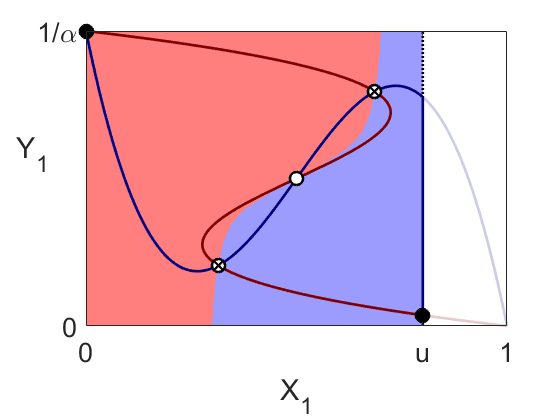}
    \caption{$u=0.8$}
    \label{fig:a6b30u0p8}
\end{subfigure}\\
\begin{subfigure}{.45\textwidth}
		\centering
		\includegraphics[width=\linewidth]{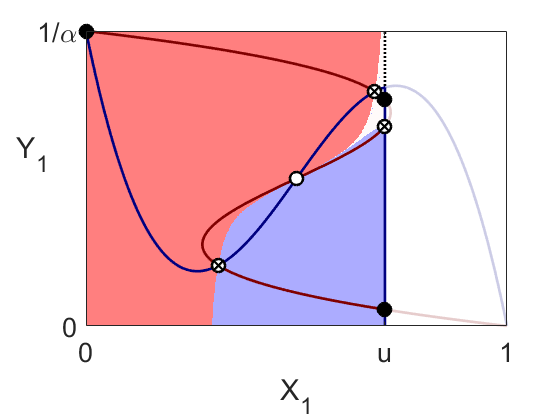}
        \caption{$u=0.71$}
        \label{fig:a6b30u0p71}
\end{subfigure}
\begin{subfigure}{.45\textwidth}
		\centering
    \includegraphics[width=\linewidth]{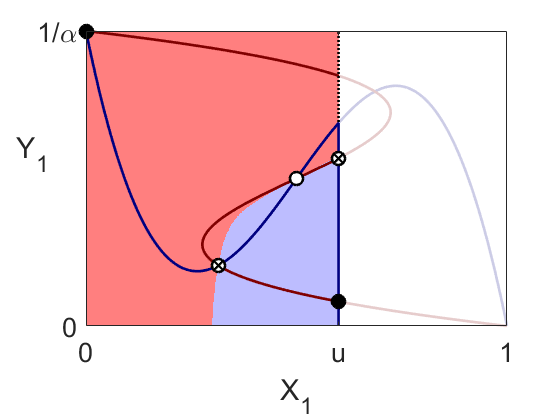}
    \caption{$u=0.6$}
    \label{fig:a6b30u0p6}
\end{subfigure}\\
\begin{subfigure}{.45\textwidth}
		\centering
		\includegraphics[width=\linewidth]{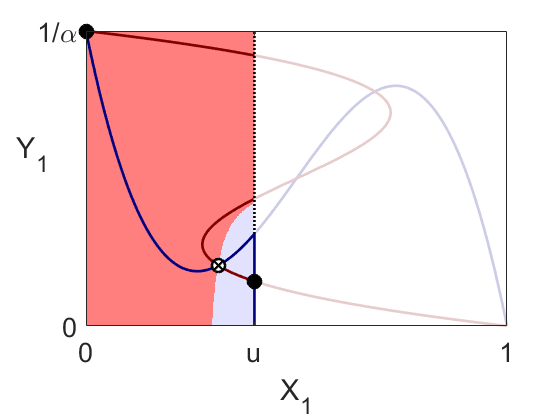}
        \caption{$u=0.4$}
        \label{fig:a6b30u0p4}
\end{subfigure}
\begin{subfigure}{.45\textwidth}
		\centering
    \includegraphics[width=\linewidth]{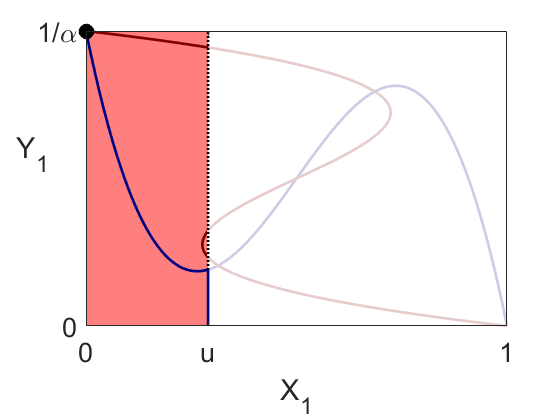}
    \caption{$u=0.29$}
    \label{fig:a6b30u0p29}
\end{subfigure}
\caption{Limiting the $X_1$ population: $(\alpha,\beta)=(6,30)$. Stable equilibria are denoted by $\CIRCLE$, unstable nodes by $\Circle$ and saddle points by $\bigotimes$.}
\label{fig:a6b30u}
\end{figure}

We demonstrate the effects of restricting the $X_1$-population in two different areas of $(\alpha,\beta)$ parameter space shown in figure~\ref{fig:limitnumbers}. First we take $(\alpha,\beta)=(6,30)$, so we are in the light shaded region of figure~\ref{pspace1}. When $u=1$, shown in figure~\ref{fig:a6b30u1p0}, we have no restriction on population. We have 5 equilibria: two stable segregated equilibria with their basins of attraction, plus three unstable integrated equilibria (a central unstable node, flanked by two saddles).  When $u=0.8$, we see in figure~\ref{fig:a6b30u0p8} that the segregated equilibrium at $(X_1,Y_1)=(1,0)$ is no longer accessible, being replaced by a (slightly) integrated stable equilibrium. The three unstable integrated equilibria still survive. A further decrease to $u=0.71$ produces two extra equilibria (figure~\ref{fig:a6b30u0p71}). One of these is a stable highly integrated equilibrium, but it has a small basin of attraction, shown in white to the left of the line $X_1=0.71$. This equilibrium only exists for $u \in [X_1^a,X_1^+]$. When $u=0.6$ (figure~\ref{fig:a6b30u0p6}), we see that we have lost two equilbria, leaving us with just two equilibria on the line $X_1=0.6$, with only the lower one being stable. 
Subsequent reduction to $u=0.4$ leads to a loss of a further two equilibria (figure~\ref{fig:a6b30u0p4}). Finally when $u=0.29$, we lose yet another pair of equilibria, leaving only the stable equilibrium at  $(X_1,Y_1)=(0,\frac{1}{\alpha})$. 

\begin{figure}[h!]
\centering
\begin{subfigure}{.45\textwidth}
		\centering
		\includegraphics[width=\linewidth]{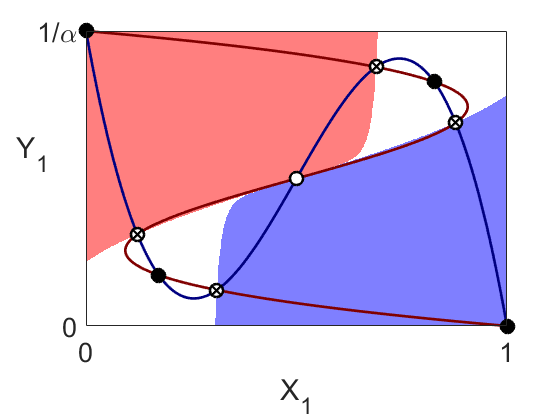}
        \caption{$u=1$}
        \label{fig:a7b49u1p0}
\end{subfigure}
\begin{subfigure}{.45\textwidth}
		\centering
    \includegraphics[width=\linewidth]{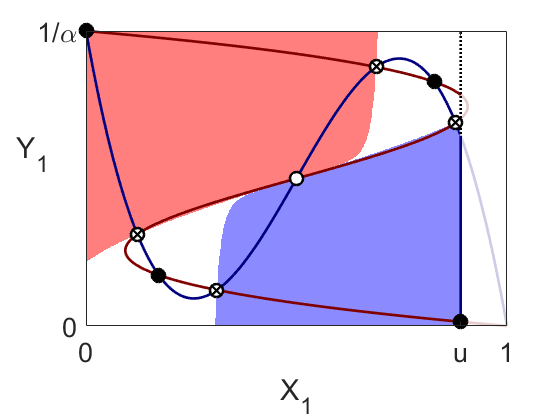}
    \caption{$u=0.89$}
    \label{fig:a7b49u0p89}
\end{subfigure}\\
\begin{subfigure}{.45\textwidth}
		\centering
		\includegraphics[width=\linewidth]{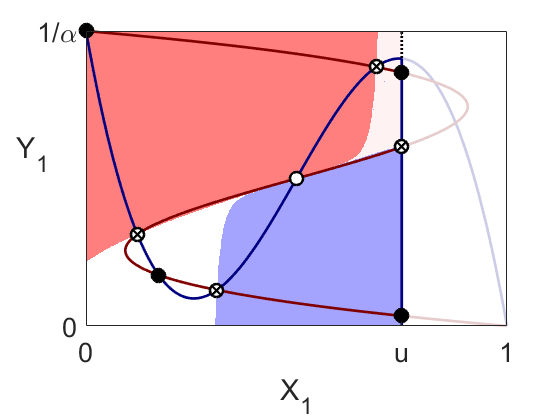}
        \caption{$u=0.75$}
        \label{fig:a7b49u0p75}
\end{subfigure}
\begin{subfigure}{.45\textwidth}
		\centering
    \includegraphics[width=\linewidth]{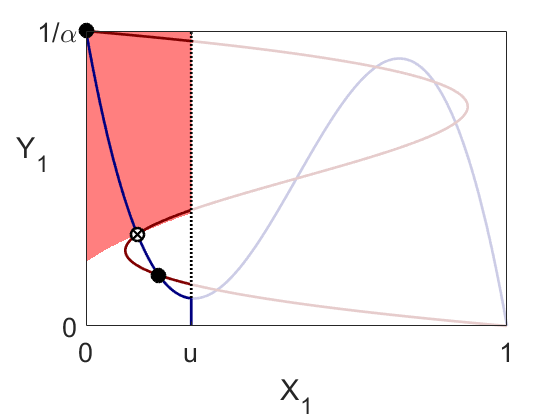}
    \caption{$u=0.25$}
    \label{fig:a7b49u0p25}
\end{subfigure}\\
\begin{subfigure}{.45\textwidth}
		\centering
		\includegraphics[width=\linewidth]{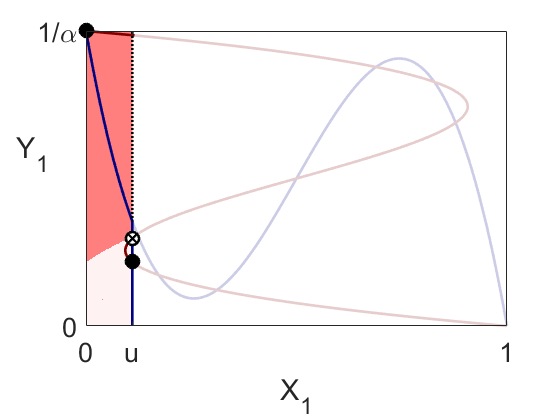}
        \caption{$u=0.11$}
        \label{fig:a7b49u0p11}
\end{subfigure}
\begin{subfigure}{.45\textwidth}
		\centering
    \includegraphics[width=\linewidth]{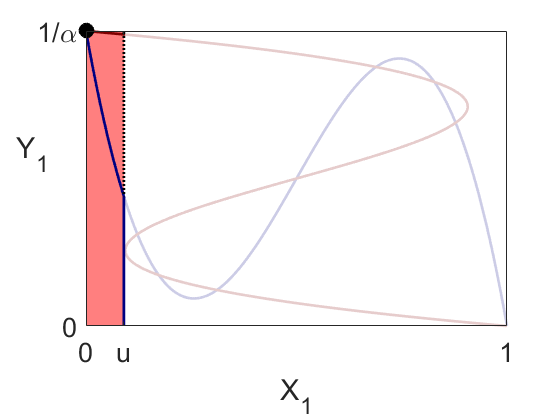}
    \caption{$u=0.09$}
    \label{fig:a7b49u0p09}
\end{subfigure}
\caption{Limiting the $X_1$ population: $(\alpha,\beta)=(7,49)$. Stable equilibria are denoted by $\CIRCLE$, unstable nodes by $\Circle$ and saddle points by $\bigotimes$.}
\label{fig:a7b49u}
\end{figure}

In our second case, we take $(\alpha,\beta)=(7,49)$, so we are in the dark shaded region of figure~\ref{pspace1}. Now we have 9 equilibria, two of which correspond to stable integration. The effect of restricting the $X_1$-population is shown in figure~\ref{fig:a7b49u}, for $u=1,\,0.89,\,0.75,\,0.25,\,0.11,\,0.09$. 

Hence it can be seen that in the two neighbourhood case, restriction of population does not always lead to new integrated equilibria, and those that are produced may only have small basins of attraction. Of possible greater concern is that population restriction can also eliminate integration.

\section{Nonlinear tolerance schedules}
\label{sec:nonlin}
In this section, we consider other ways in which the linear tolerance schedule can be modified to produce integrated populations. We illustrate phenomena that can occur when the tolerance schedules are nonlinear, using the original equations \eqref{eq:orig} in their unscaled form, without restricting the population. We focus on the case $a=2,\,b=10,\,k=1$, that is $(\alpha,\beta)=(2,20)$. This is the simplest case when the tolerance schedules are linear, corresponding to dynamics in the white region of figure~\ref{pspace1}, where we have two stable segregated equilibria and one unstable integrated equilibria. 

\begin{figure}[h!]
\centering
		\begin{subfigure}{.45\textwidth}
		\includegraphics[width=\linewidth]{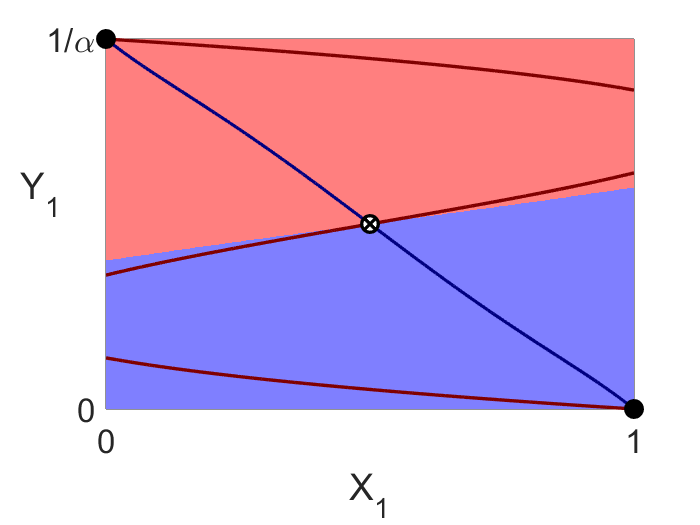}
		\caption{$R_{X_1}=RE_{X_1}^4$; $R_{Y_1}= 10(1-Y_1)$.}
		\label{fig:explin}
		\end{subfigure}
		\begin{subfigure}{.45\textwidth}
		\includegraphics[width=\linewidth]{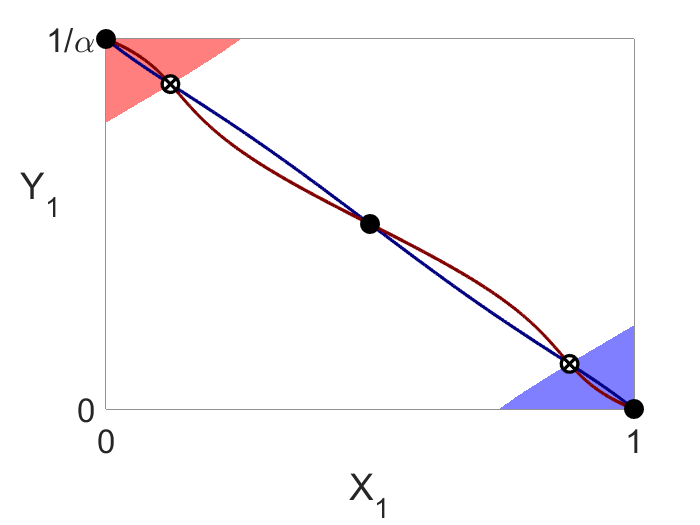}
		\caption{$R_{X_1}=RE_{X_1}^4$; $R_{Y_1}=RE_{Y_1}^4$.}
		\label{fig:twoexp}
		\end{subfigure}\\
        \begin{subfigure}{.45\textwidth}
        \includegraphics[width=\linewidth]{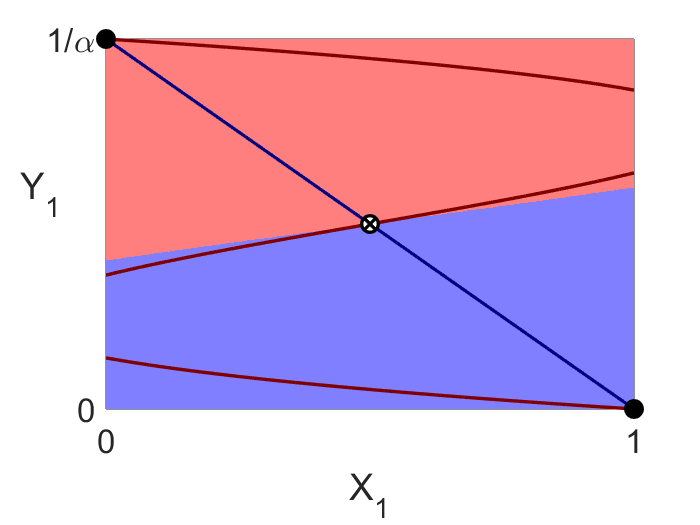}
        \caption{$R_{X_1}=RQ_{X_1}^2$; $R_{Y_1}= 10(1-Y_1)$.}
        \label{fig:quadlin}
        \end{subfigure}
        \begin{subfigure}{.45\textwidth}
        \includegraphics[width=\linewidth]{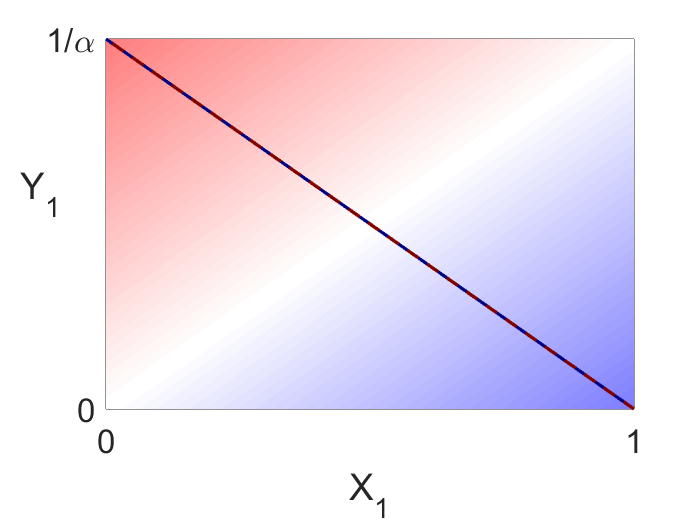}
        \caption{$R_{X_1}=RQ_{X_1}^2$; $R_{Y_1}=RQ_{Y_1}^2$.}
        \label{fig:twoquad}
        \end{subfigure}
        \begin{subfigure}{.45\textwidth}
        \includegraphics[width=\linewidth]{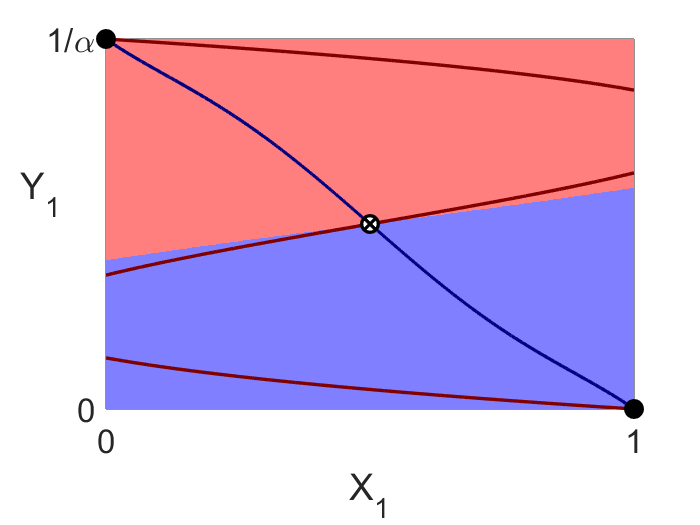}
        \caption{$R_{X_1}=RQ_{X_1}^3$; $R_{Y_1}= 10(1-Y_1)$.}
        \label{fig:cublin}
        \end{subfigure}
        \begin{subfigure}{.45\textwidth}
        \includegraphics[width=\linewidth]{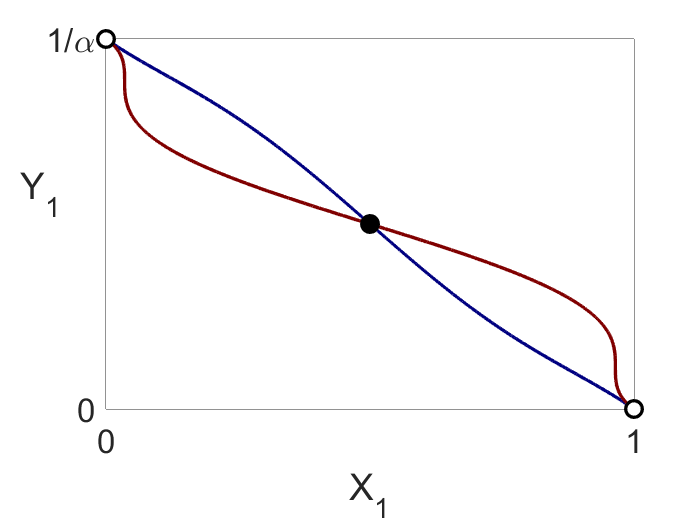}
        \caption{$R_{X_1}=RQ_{X_1}^3$; $R_{Y_1}=RQ_{Y_1}^3$.
        \label{fig:twocub}}
        \end{subfigure}

\caption{Nullclines, fixed points and basins of attraction for some candidate nonlinear tolerance schedules. Stable equilibria are denoted by $\CIRCLE$, unstable nodes by $\Circle$ and saddle points by $\bigotimes$.} 
\label{rq3phase}
\end{figure}

In our first example, let us replace the linear $X_1$-population tolerance schedule \eqref{eq:tol} with an exponential tolerance schedule \cite{Haw2018}, of the form
\begin{align}
\label{eq:REX4}
R_{X_1}(X_1)=&RE_{X_1}^4(X_1)\equiv\frac{2}{1-e^{-4}}\left[e^{-4 X_1}-e^{-4}\right],
\end{align}
where $RE_{X_1}^4(0)=2,\,RE_{X_1}^4(1)=0$. The $Y_1$-population has the linear tolerance schedule $R_{Y_1}(Y_1) \equiv 10(1-Y_1)$. The results shown in figure~\ref{fig:explin} are similar to the linear case. 

Keeping $R_{X_1}(X_1)=RE_{X_1}^4(X_1)$, we now replace the linear tolerance schedule of the $Y_1$-population with an exponential tolerance schedule of the form
\begin{align}
\label{eq:REY4}
R_{Y_1}(Y_1)=&RE_{Y_1}^4(Y_1)\equiv\frac{10}{1-e^{-4}}\left[e^{-4 Y_1}-e^{-4}\right],
\end{align}
where $RE_{Y_1}^4(0)=10,\,RE_{Y_1}^4(1)=0$.
In this case, the central fixed point - a fully integrated population in both neighbourhoods - is now stable, with a large basin of attraction (figure~\ref{fig:twoexp}). However, near the saddle points, a small change in initial conditions can lead to a stable segregated population.



Next, consider polynomial tolerance schedules \cite{Haw2018}, of the form
\begin{align}
R_{X_1}(X_1)=&RQ_{X_1}^p(X_1)\equiv a(1-X_1)^p\label{eq:Xpol}\\ 
R_{Y_1}(Y_1)=&RQ_{Y_1}^p(Y_1)\equiv b(1-kY_1)^p,\label{eq:Ypol}
\end{align}
where $p\in \mathcal{Z}^+$ and $a=2,\,b=10,\,k=1$, that is $(\alpha,\beta)=(2,20)$, as before. When $p=1$, we have the linear tolerance schedule case. When $p=2$, the corresponding nullcline is a straight line in $(X_1,Y_1)$ phase space. In figure~\ref{fig:quadlin}, we set $R_{X_1}=RQ_{X_1}^2$, given by \eqref{eq:Xpol} and $R_{Y_1}(Y_1)= 10(1-Y_1)$, that is, $p=2$ for $X_1$ and $p=1$ for $Y_1$. The results are similar to both the linear case and to figure~\ref{fig:explin}. However when we set we set $R_{X_1}=RQ_{X_1}^2$, and $R_{Y_1}=RQ_{Y_1}^2$, we have an open set of equilibria, given by the line $\alpha Y_1=1-X_1$, shown in figure~\ref{fig:twoquad}. Any desired population mixture can be obtained simply by the correct choice of initial conditions. However it is clear that this outcome is not robust - any small change in $p$ will lead to a different outcome.

Finally, in figure~\ref{fig:cublin}, we set $R_{X_1}=RQ_{X_1}^3$, and $R_{Y_1}(Y_1)= 10(1-Y_1)$. The results are similar to both the linear case and to figures~\ref{fig:explin} and~\ref{fig:quadlin}. However when we set we set $R_{X_1}=RQ_{X_1}^3$, and $R_{Y_1}=RQ_{Y_1}^3$, the central fixed point is stable and its basin of attraction covers the whole of phase space, shown in figure \ref{fig:twocub}. 

$RQ_X^3$ and $RQ_Y^3$ represent globally \textit{less} tolerant populations than the equivalent linear cases. We can interpret the stable fixed point in figure~\ref{fig:twocub} phenomenon as follows: low tolerance constraints mean that both types are trying to leave both neighbourhoods simultaneously. The stable equilibrium therefore represents a best case scenario, rather than a state in which all tolerance demands are satisfied. 

So overall we can conclude that modification of the linear tolerance schedule of only one population may not be enough to induce integration. Instead, both populations may need to modify their tolerance schedules to become integrated.

\section{Discussion}
\label{sec:disc}

Schelling \cite{Schelling2006} states that ``\dots [in the BNM] \ldots an important phenomenon can be that a too-tolerant majority can overwhelm a minority and bring about segregation". We have that $$\frac{\beta}{b}=a=\frac{\alpha}{k}.$$ So $\beta =\frac{b}{k}\alpha$. Hence lines through the origin of $(\alpha, \beta)$ parameter space correspond to increasing $a$, the upper tolerance limit of the majority $X$-population.
We show two of these lines in Figure~\ref{fig:limitnumbers}, for $\frac{b}{k}=2,8$ corresponding to the case when middle branch of nullcline \eqref{sub2a} both exists and lies wholly within the allowed phase space. As we can see, asymptotic to that range is the area of existence of stable integration. Outside that range, even if $a$ is very big, we can not get stable integration, thus quantifying Schelling's \cite{Schelling2006} statement. 

So far, we have only considered equilibria of our governing Schelling dynamical system \eqref{eq:linunscaled}. Do these equations have periodic solutions? They have been observed in discrete time ``two rooms" models of segregation. But these oscillations appear to be neutrally stable and consist of population swings in both rooms. Periodic solutions have also been observed \cite{Gramlich2016,Jansen2001} in Lotka-Volterra predator-prey models in two habitats (or patches). But there the dynamics is substantially different.

In our case, we have not found any Hopf bifurcations in our calculations and extensive numerical simulations have not produced any limit cycles. So if they exist, they are most likely unstable (or have a very small basin of attraction). Since \eqref{eq:linunscaled} is a planar system,  Dulac's criterion \cite{JordanSmithhBook} could be used to show that \eqref{eq:linunscaled} does not have limit cycles. But up to now, we have been unable to find the correct Dulac function. So the existence and stability of periodic solutions to \eqref{eq:linunscaled} must remain an open question.

We can use our results to consider the effects of variation in parameters $\alpha$ and $\beta$. In particular we are interested in what might happen as the minority $Y$-population grows. Provided the variation is slow enough to be considered quasi-static, we can simply move around parameter space. A key parameter is $\alpha \equiv ak$. If we fix $a$, the maximum tolerance of the $X$-population, and then decrease $k$, the $Y$-population grows as $\alpha$ decreases. Then for fixed combined tolerance parameter $\beta \equiv ab >36$ in figure~\ref{pspace1}, we see that integration is only a transient phase as we enter and then leave the dark shaded region. So as $k$ decreases, we need to ensure that $\alpha$ stays fixed, and that can only happen if $a$ {\it increases}. In other words, when a minority grows, stable integration is only possible if the majority population increases its own tolerance as well. This runs counter to the populist idea that a growing minority should integrate more into the majority to be accepted.

In \cite{Haw2018}, we considered the case where the two populations could live either in one neighbourhood or remove themselves to a place ``where colour does not matter" \cite{Schelling1971}. Suppose now that this place changes in such a way that colour does matter. This could happen for example by the creation or removal of borders or as the result of a change in government. If there is an integrated population in the single neighbourhood, will it remain integrated after the change to two neighbourhoods?

\begin{figure}[h!]
	\centering
	    \includegraphics[width=.7\textwidth]{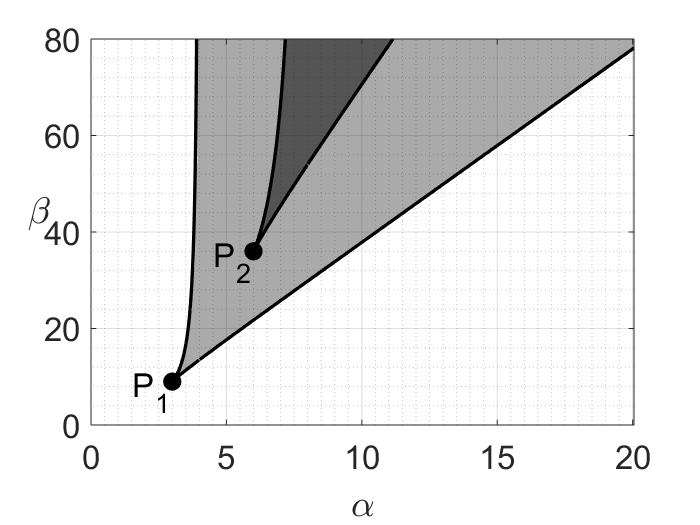}
	    \caption{Comparison of integrated population parameters for two-neighbourhoods with the single-neighbourhood problem \cite{Haw2018}. Curves with apex $P_1=(3,9)$ are $\beta=\beta_{\pm}$ from \cite{Haw2018}. Curves with apex $P_2=(6,36)$ are $\beta=\beta_{\pm}$ from \eqref{eq:3real} above.}
	    \label{pspacecompare}
\end{figure}

If we overlay part of figure~\ref{pspace1} with part of figure 2 of \cite{Haw2018}, we obtain figure~\ref{pspacecompare}. We can now make the following observation: in the light shaded region of figure~\ref{pspacecompare} with apex $P_1$, a single neighbourhood can have a stable, integrated population. But, for the two neighbourhood case, a stable integrated population is possible only in the dark shaded region with apex $P_2$. 

We then arrive at the remarkable conclusion that, if the minority population is such that its values of $\alpha, \beta$ lie in the light shaded region, a re-organisation of neighbourhoods can lead to the loss of the stable integrated population, without any change in the numbers or attitudes of {\it either} population. 
Put another way, some types of minority population in two neighbourhoods can only achieve integration by creating one neighbourhood and a place where type does not matter, or after a perturbation of the system (tipping) into the basin of attraction of an integrated equilibrium. 

The two curves shown in figure~\ref{pspacecompare} appear similar. In fact, a simple mapping takes one curve into the other.  Apex $P_2$ has coordinates $(\alpha,\beta) = (6,36)$, whereas at apex $P_1$, $(\alpha,\beta) = (3,9)$. If we set $(\alpha,\beta)_{2-neighbourhood}=(2\alpha,4\beta)_{1-neighbourhood}$ in \eqref{eq:disc}, then we recover equation (9) of \cite{Haw2018}. These equations give curves in parameter space where the discriminant of cubic is zero. But it can be shown that the cubic equations from which they originate are completely different and that the governing dynamical systems cannot be mapped to one another.

\begin{figure}[h!]
	\centering
	\begin{tikzpicture}[scale=1,transform shape,ultra thick,main node/.style={rectangle,draw,font=\sffamily\large}]
    \node[main node,fill=davy!30,minimum width=25pt] (1) at (-4,0) {neighbourhood $1$ $(X_1,Y_1)$};
	\node[main node,fill=davy!30,minimum width=25pt] (2) at (4,0) {neighbourhood $2$ $(X_2,Y_2)$};
	\node[main node,fill=davy!30,minimum width=25pt,style=circle] (3) at (0,2) {$X_3$};
    \node[main node,fill=davy!30,minimum width=25pt,style=circle] (4) at (0,-2) {$Y_3$};
	\path[every node/.style={font=\sffamily\large}]
	(1) edge [line width=1.5pt,->,bend left] node[auto] {$-\dot X_1$} (3)
	(3) edge [dotted,line width=1.5pt,->] node[auto] {$\dot X_1$} (1.north east)
    (1) edge [line width=1.5pt,->,bend right] node[below] {$-\dot Y_1$} (4)
	(4) edge [dotted,line width=1.5pt,->] node[above] {$\dot Y_1$} (1.south east)
    (2) edge [line width=1.5pt,->,bend right] node[above] {$-\dot X_2$} (3)
	(3) edge [dotted,line width=1.5pt,->] node[below] {$\dot X_2$} (2.north west)
    (2) edge [line width=1.5pt,->,bend left] node[auto] {$-\dot Y_2$} (4)
	(4) edge [dotted,line width=1.5pt,->] node[auto] {$\dot Y_2$} (2.south west);
\end{tikzpicture}
\caption{Dynamics of the two neighbourhoods problem with reservoirs of population. All arrows represent population flow when the corresponding quantities are positive. Dotted arrows have the additional constraint that $X_3>0$ (top part of schematic) or $Y_3>0$ (bottom part of schematic).}
\label{fig:schem}
\end{figure}
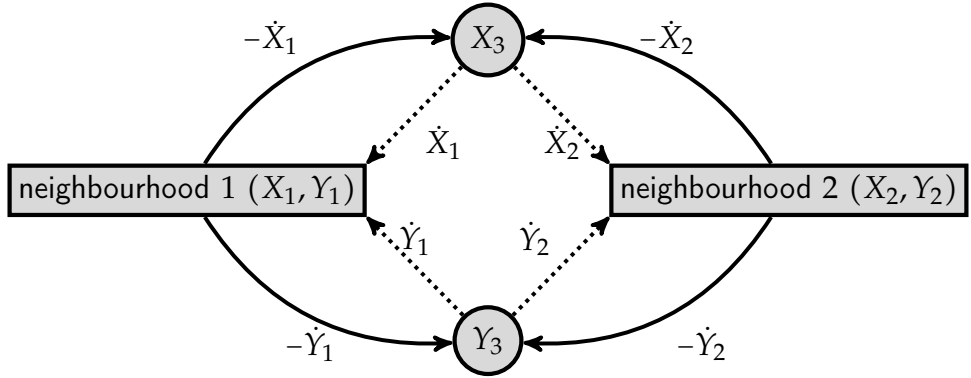

We can extend the two-neighbourhood problem, by introducing an option to be in neither neighbourhood (figure \ref{fig:schem}). Let $X_3$ and $Y_3$ denote the $X$- and $Y$-populations present in neither neighbourhood. These reservoirs are segregated. Thus $X_3=1-X_1-X_2$ and $Y_3=\frac{1}{\alpha}-Y_1-Y_2$. Neighbourhoods 1 and 2 are occupied upon demand as follows.  People enter a neighbourhood when the tolerance in that neighbourhood means that they would stay (and of course people leave a neighbourhood when the tolerance there means they should leave). To move between neighbourhoods, population members must go through either $X_3$ or $Y_3$ (in this version of the problem, there is no direct movement between  neighbourhoods 1 and 2).

If $f_1(X_i,Y_i)=X_i(X_iR_{X_i}(X_i)-Y_i), \, i=1,2$ and $g_1(X_i,Y_i)=Y_i(Y_iR_{Y_i}(Y_i)-Y_i), \, i=1,2$, then
\begin{eqnarray}
\frac{dX_1}{dt}=
\begin{cases} 
      f_1(X_1,Y_1) & f_1(X_1,Y_1)\leq 0\quad\text{OR}\quad X_3>0\\
      \max(0,-f_2) & f_1(X_1,Y_1)>0\quad\text{AND}\quad X_3=0 \label{eq:twopopresX}
\end{cases}\\
\frac{dY_1}{dt}=
\begin{cases} 
      g_1(X_1,Y_1) & g_1(X_1,Y_1)\leq 0\quad\text{OR}\quad Y_3>0\\
      \max(0,-g_2) & g_1(X_1,Y_1)>0\quad\text{AND}\quad Y_3=0, \label{eq:twopopresY}
\end{cases}
\end{eqnarray}
which is equivalent to the single-neighbourhood case with limiting numbers and $u=X_1+X_3,\ v=Y_1+Y_3$. We summarise the dynamics in table~\ref{table:case2cases}. 

\begin{table}[h]
\centering
\begin{tabular}{c|c|c||c|c|c|}
& $X_3>0$ & $X_3=0$ & & $Y_3>0$ & $Y_3=0$\\
\hline
$f_1\leq 0$ & $f_1$ & $f_1$ & $g_1\leq 0$ & $g_1$ & $g_1$\\
\hline
$f_1>0$ & $f_1$ & $0$ & $g_1>0$ & $g_1$ & $0$\\
\hline
\hline
$f_2\leq 0$ & $f_2$ & $f_2$ & $g_2\leq 0$ & $g_2$ & $g_2$\\
\hline
$f_2>0$ & $f_2$ & $0$ & $g_2>0$ & $g_2$ & $0$\\
\hline
\end{tabular}
\caption{Summary of dynamics for two neighbourhoods with population reservoirs, given in \eqref{eq:twopopresX} and \eqref{eq:twopopresY}. First two columns: $\frac{dX_1}{dt}$. Second two columns: $\frac{dY_1}{dt}$.}
\label{table:case2cases}
\end{table}

\begin{figure}[h!]
	\centering
	\begin{subfigure}{.49\textwidth}
		\centering
	    \includegraphics[width=\textwidth]{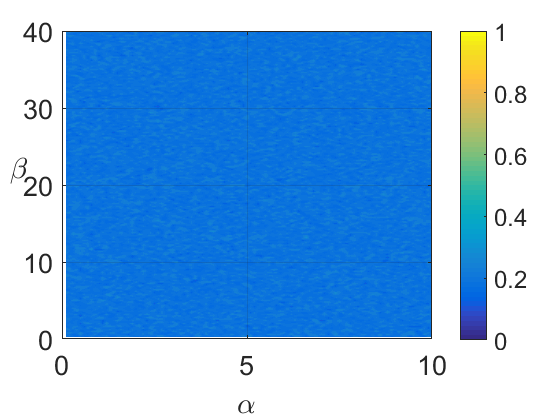}
	    \caption{Two-neighbourhood model with reservoirs.\newline}
	\end{subfigure}
	\begin{subfigure}{.49\textwidth}
		\centering
	    \includegraphics[width=\textwidth]{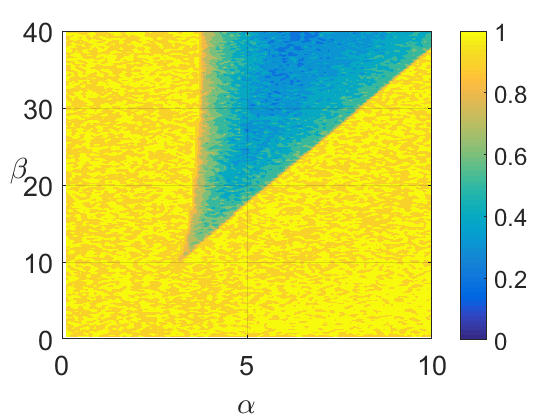}
	    \caption{$2$ independent single neighbourhood models without conservation of population.}
	\end{subfigure}\\
	\begin{subfigure}{.49\textwidth}
		\centering
		\includegraphics[width=\linewidth]{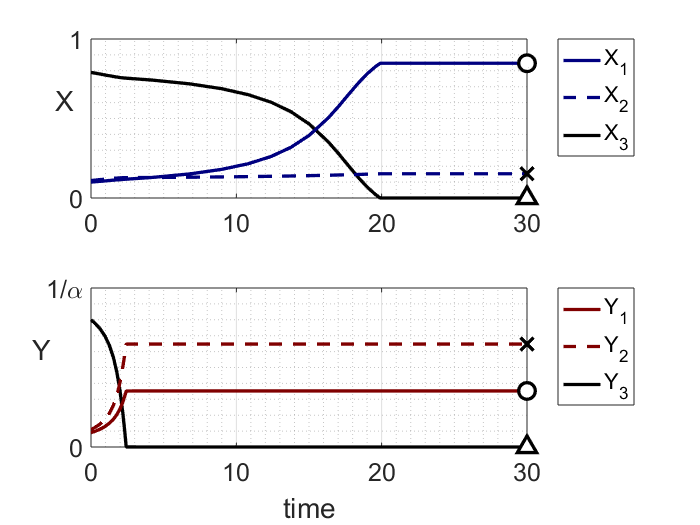}
		\caption{Example trajectories (two-neighbourhood model).}
	\end{subfigure}
	\begin{subfigure}{.49\textwidth}
		\centering
		\includegraphics[width=\linewidth]{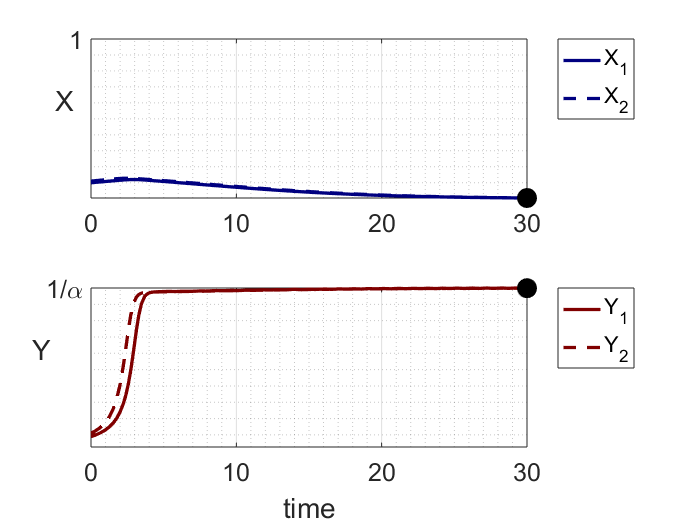}
		\caption{Example trajectories (single neighbourhood models).}
		\end{subfigure}
	\caption{For each $(\alpha,\beta)$, we simulated $20$ sets of initial conditions for $t=0$ to $t=200$, calculated dissimilarity $\mathcal{D}$ for each and plotted the mean $\mathcal{D}$ in parameter space ($\mathcal{D}=1$ indicates complete segregation, $\mathcal{D}=0$ indicates an even distribution between neighbourhoods). Note that two empty neighbourhoods yields $\mathcal{D}=0$. Results for two neighbourhoods with reservoirs are shown in (a), and two independent simulations without conservation of total population are shown in (b). Example trajectories from (a) and (b) are given in (c) and (d) respectively, with $\alpha=6,\beta=30,X_1(0)=0.1,X_2(0)=0.11,Y_1(0)=0.09,Y_2(0)=0.11$.}
	\label{fig:reserve}
\end{figure}

Figure~\ref{fig:reserve} shows the results of simulating this system. For each $(\alpha,\beta)$, we simulate $20$ randomly selected initial conditions, and compute the index of dissimilarity $\mathcal{D}$ at equilibrium. The conventional definition of dissimilarity \cite{Massey1988} is $\mathcal{D}:=\frac 12[|X_1-\alpha Y_1|+|X_2-\alpha Y_2|]\in[0,1]$, and corresponds to the proportion of the minority population that would have to relocate in order to yield a uniform distribution of both types across all neighbourhoods. One typical criticism of the index of dissimilarity is its sensitivity to neighbourhood boundaries. Since boundaries are an inherent property of any BNM, this index is a natural tool to employ in our study. 

Our colour scale for basins of attraction in cases I-IV uses the modified measure $\bar{\mathcal{D}}:=X_1-\alpha Y_1\in[-1,1]$ applied to the corresponding stable equilibrium, in order to distinguish between $X_1$- and $Y_1$-dominance. A white basin of attraction denotes an even distribution of types in both neighbourhoods. In figure~\ref{fig:reserve}, we return to the conventional $\mathcal{D}\in [0,1]$ as this yields a single measure for the whole system, rather than just one neighbourhoods. 

We repeat this procedure for systems with two neighbourhoods without conservation of population, namely two decoupled, single-neighbourhood models, as in \cite{Haw2018}. 
In the former case, the constraint imposed by conservation typically results in stable integrated states. 
In the latter case, the colour scale clearly shows the relative area of basin of attraction of stable segregated states ($\mathcal{D}=1$), weighted by dissimilarity of the stable integrated states. 

\section{Conclusion}
\label{sec:conclusion}
We have considered Schelling's BNM for the case of two neighbourhoods, with and without population reservoirs. In the latter case, we presented the governing dynamical systems in \eqref{eq:case1X} and \eqref{eq:case1Y}. For the case of identical $(X,Y)$ linear tolerance schedules (case I), we have carried out an extensive analysis, showing both the existence and stability of integration. We have shown that such an outcome requires a small minority, in the presence of a highly tolerant majority. Similar results were obtained when the linear tolerances differ between neighbourhoods (cases II, III, IV).

If one or the other population is restricted, by removing the most intolerant individuals, we have shown that new stable population mixtures can be created. But they may only exist in narrow regions of parameter space, with small basins of attraction. In some cases, existing stable integrated populations can even be destroyed by this process. We also considered how different nonlinear tolerance schedules can affect our results.

Our results shed light on some popular notions of integration. So if a minority grows slowly, existing stable integration will be destroyed unless the majority population becomes more tolerant (and yet they may feel less tolerant as a result of the increase in the minority). Any increase in tolerance by the minority will have little effect. 

Similarly the transition from one neighbourhood to two neighbourhoods (or vice versa) is not necessarily straightforward. A well-integrated single neighbourhood can become segregated after such a transition, without any change in the tolerance of either population.

Finally, when considering the case of $2$ neighbourhoods connected by population reserves, we see that integrated stable states exist for low values of tolerance parameters as a result of competition for finite resources. The notion of trade-off between tolerance demands and external constraints, namely neighbourhood structures and finite population reserves, is crucial to understanding the dynamics of segregation. We invite readers from socio-economic disciplines to offer insight into interpretation of our results, and to aid the construction of more complex model variants that better describe the flows of population in real urban areas. 
\newpage
\bibliographystyle{plain}

\end{document}